\documentclass[twocolumns,traditabstract,longauth]{aa} 
\usepackage{graphicx}
\usepackage{txfonts}
\usepackage{url}
\usepackage[breaklinks, colorlinks, citecolor=blue]{hyperref} 
\usepackage{upgreek}
\usepackage{natbib}
\bibpunct{(}{)}{;}{a}{}{,} 
\usepackage{amssymb}%
\usepackage{rotating}
\usepackage{longtable}
\usepackage{supertabular}
\usepackage{lscape}
\usepackage{epsfig}
\usepackage{enumerate}
\usepackage{breakurl}

\def\setsymbol#1#2{\expandafter\def\csname #1\endcsname{#2}}
\def\getsymbol#1{\csname #1\endcsname}

\def\Planck{\textit{Planck}}

\newbox\tablebox    \newdimen\tablewidth
\def\leaderfil{\leaders\hbox to 5pt{\hss.\hss}\hfil}
\def\endPlancktable{\tablewidth=\columnwidth
     $$\hss\copy\tablebox\hss$$
     \vskip-\lastskip\vskip -2pt}
\def\endPlancktablewide{\tablewidth=\textwidth
     $$\hss\copy\tablebox\hss$$
     \vskip-\lastskip\vskip -2pt}
\def\tablenote#1 #2\par{\begingroup \parindent=0.8em
     \abovedisplayshortskip=0pt\belowdisplayshortskip=0pt
     \noindent
     $$\hss\vbox{\hsize\tablewidth \hangindent=\parindent \hangafter=1 \noindent
     \hbox to \parindent{$^#1$\hss}\strut#2\strut\par}\hss$$
     \endgroup}
\def\doubleline{\vskip 3pt\hrule \vskip 1.5pt \hrule \vskip 5pt}

\def\L2{\ifmmode L_2\else $L_2$\fi}

\def\DeltaT{\ifmmode \Delta T\else $\Delta T$\fi}
\def\deltat{\ifmmode \Delta t\else $\Delta t$\fi}
\def\fknee{\ifmmode f_{\rm knee}\else $f_{\rm knee}$\fi}
\def\Fmax{\ifmmode F_{\rm max}\else $F_{\rm max}$\fi}
\def\solar{\ifmmode{\rm M}_{\mathord\odot}\else${\rm M}_{\mathord\odot}$\fi}
\def\Msolar{\ifmmode{\rm M}_{\mathord\odot}\else${\rm M}_{\mathord\odot}$\fi}
\def\Lsolar{\ifmmode{\rm L}_{\mathord\odot}\else${\rm L}_{\mathord\odot}$\fi}

\def\inv{\ifmmode^{-1}\else$^{-1}$\fi}
\def\mo{\ifmmode^{-1}\else$^{-1}$\fi}
\def\sup#1{\ifmmode ^{\rm #1}\else $^{\rm #1}$\fi}
\def\expo#1{\ifmmode \times 10^{#1}\else $\times 10^{#1}$\fi}
\def\,{\thinspace}
\def\lsim{\mathrel{\raise .4ex\hbox{\rlap{$<$}\lower 1.2ex\hbox{$\sim$}}}}
\def\gsim{\mathrel{\raise .4ex\hbox{\rlap{$>$}\lower 1.2ex\hbox{$\sim$}}}}

\def\simprop{\mathrel{\raise .4ex\hbox{\rlap{$\propto$}\lower 
1.2ex\hbox{$\sim$}}}}
\def\deg{\ifmmode^\circ\else$^\circ$\fi}
\def\pdeg{\ifmmode $\setbox0=\hbox{$^{\circ}$}\rlap{\hskip.11\wd0 .}$^{\circ}
           \else \setbox0=\hbox{$^{\circ}$}\rlap{\hskip.11\wd0 .}$^{\circ}$\fi}
\def\arcs{\ifmmode {^{\scriptstyle\prime\prime}}
           \else $^{\scriptstyle\prime\prime}$\fi}
\def\arcm{\ifmmode {^{\scriptstyle\prime}}
           \else $^{\scriptstyle\prime}$\fi}
\newdimen\sa  \newdimen\sb
\def\parcs{\sa=.07em \sb=.03em
      \ifmmode \hbox{\rlap{.}}^{\scriptstyle\prime\kern 
-\sb\prime}\hbox{\kern -\sa}
      \else \rlap{.}$^{\scriptstyle\prime\kern -\sb\prime}$\kern -\sa\fi}
\def\parcm{\sa=.08em \sb=.03em
      \ifmmode \hbox{\rlap{.}\kern\sa}^{\scriptstyle\prime}\hbox{\kern-\sb}
      \else \rlap{.}\kern\sa$^{\scriptstyle\prime}$\kern-\sb\fi}
\def\ra[#1 #2 #3.#4]{#1\sup{h}#2\sup{m}#3\sup{s}\llap.#4}
\def\dec[#1 #2 #3.#4]{#1\deg#2\arcm#3\arcs\llap.#4}
\def\deco[#1 #2 #3]{#1\deg#2\arcm#3\arcs}
\def\rra[#1 #2]{#1\sup{h}#2\sup{m}}

\def\dots{\relax\ifmmode \ldots\else $\ldots$\fi}
\def\WHzsr{\ifmmode $W\,Hz\mo\,sr\mo$\else W\,Hz\mo\,sr\mo\fi}
\def\mHz{\ifmmode $\,mHz$\else \,mHz\fi}
\def\GHz{\ifmmode $\,GHz$\else \,GHz\fi}
\def\mKs{\ifmmode $\,mK\,s$^{1/2}\else \,mK\,s$^{1/2}$\fi}
\def\muKs{\ifmmode \,\mu$K\,s$^{1/2}\else \,$\mu$K\,s$^{1/2}$\fi}
\def\muKRJs{\ifmmode \,\mu$K$_{\rm RJ}$\,s$^{1/2}\else \,$\mu$K$_{\rm 
RJ}$\,s$^{1/2}$\fi}
\def\muKHz{\ifmmode \,\mu$K\,Hz$^{-1/2}\else \,$\mu$K\,Hz$^{-1/2}$\fi}
\def\MJysr{\ifmmode \,$MJy\,sr\mo$\else \,MJy\,sr\mo\fi}
\def\MJysrmK{\ifmmode \,$MJy\,sr\mo$\,mK$_{\rm CMB}\mo\else 
\,MJy\,sr\mo\,mK$_{\rm CMB}\mo$\fi}
\def\microns{\ifmmode \,\mu$m$\else \,$\mu$m\fi}

\def\muK{\ifmmode \,\mu$K$\else \,$\mu$\hbox{K}\fi}
\def\microK{\ifmmode \,\mu$K$\else \,$\mu$\hbox{K}\fi}
\def\muW{\ifmmode \,\mu$W$\else \,$\mu$\hbox{W}\fi}
\def\kms{\ifmmode $\,km\,s$^{-1}\else \,km\,s$^{-1}$\fi}
\def\kmsMpc{\ifmmode $\,\kms\,Mpc\mo$\else \,\kms\,Mpc\mo\fi}
\setsymbol{LFI:center:frequency:70GHz:units}{70.3\,GHz}
\setsymbol{LFI:center:frequency:44GHz:units}{44.1\,GHz}
\setsymbol{LFI:center:frequency:30GHz:units}{28.5\,GHz}

\setsymbol{LFI:center:frequency:70GHz}{70.3}
\setsymbol{LFI:center:frequency:44GHz}{44.1}
\setsymbol{LFI:center:frequency:30GHz}{28.5}

\setsymbol{LFI:center:frequency:LFI18:Rad:M:units}{71.7\GHz}
\setsymbol{LFI:center:frequency:LFI19:Rad:M:units}{67.5\GHz}
\setsymbol{LFI:center:frequency:LFI20:Rad:M:units}{69.2\GHz}
\setsymbol{LFI:center:frequency:LFI21:Rad:M:units}{70.4\GHz}
\setsymbol{LFI:center:frequency:LFI22:Rad:M:units}{71.5\GHz}
\setsymbol{LFI:center:frequency:LFI23:Rad:M:units}{70.8\GHz}
\setsymbol{LFI:center:frequency:LFI24:Rad:M:units}{44.4\GHz}
\setsymbol{LFI:center:frequency:LFI25:Rad:M:units}{44.0\GHz}
\setsymbol{LFI:center:frequency:LFI26:Rad:M:units}{43.9\GHz}
\setsymbol{LFI:center:frequency:LFI27:Rad:M:units}{28.3\GHz}
\setsymbol{LFI:center:frequency:LFI28:Rad:M:units}{28.8\GHz}
\setsymbol{LFI:center:frequency:LFI18:Rad:S:units}{70.1\GHz}
\setsymbol{LFI:center:frequency:LFI19:Rad:S:units}{69.6\GHz}
\setsymbol{LFI:center:frequency:LFI20:Rad:S:units}{69.5\GHz}
\setsymbol{LFI:center:frequency:LFI21:Rad:S:units}{69.5\GHz}
\setsymbol{LFI:center:frequency:LFI22:Rad:S:units}{72.8\GHz}
\setsymbol{LFI:center:frequency:LFI23:Rad:S:units}{71.3\GHz}
\setsymbol{LFI:center:frequency:LFI24:Rad:S:units}{44.1\GHz}
\setsymbol{LFI:center:frequency:LFI25:Rad:S:units}{44.1\GHz}
\setsymbol{LFI:center:frequency:LFI26:Rad:S:units}{44.1\GHz}
\setsymbol{LFI:center:frequency:LFI27:Rad:S:units}{28.5\GHz}
\setsymbol{LFI:center:frequency:LFI28:Rad:S:units}{28.2\GHz}

\setsymbol{LFI:center:frequency:LFI18:Rad:M}{71.7}
\setsymbol{LFI:center:frequency:LFI19:Rad:M}{67.5}
\setsymbol{LFI:center:frequency:LFI20:Rad:M}{69.2}
\setsymbol{LFI:center:frequency:LFI21:Rad:M}{70.4}
\setsymbol{LFI:center:frequency:LFI22:Rad:M}{71.5}
\setsymbol{LFI:center:frequency:LFI23:Rad:M}{70.8}
\setsymbol{LFI:center:frequency:LFI24:Rad:M}{44.4}
\setsymbol{LFI:center:frequency:LFI25:Rad:M}{44.0}
\setsymbol{LFI:center:frequency:LFI26:Rad:M}{43.9}
\setsymbol{LFI:center:frequency:LFI27:Rad:M}{28.3}
\setsymbol{LFI:center:frequency:LFI28:Rad:M}{28.8}
\setsymbol{LFI:center:frequency:LFI18:Rad:S}{70.1}
\setsymbol{LFI:center:frequency:LFI19:Rad:S}{69.6}
\setsymbol{LFI:center:frequency:LFI20:Rad:S}{69.5}
\setsymbol{LFI:center:frequency:LFI21:Rad:S}{69.5}
\setsymbol{LFI:center:frequency:LFI22:Rad:S}{72.8}
\setsymbol{LFI:center:frequency:LFI23:Rad:S}{71.3}
\setsymbol{LFI:center:frequency:LFI24:Rad:S}{44.1}
\setsymbol{LFI:center:frequency:LFI25:Rad:S}{44.1}
\setsymbol{LFI:center:frequency:LFI26:Rad:S}{44.1}
\setsymbol{LFI:center:frequency:LFI27:Rad:S}{28.5}
\setsymbol{LFI:center:frequency:LFI28:Rad:S}{28.2}

\setsymbol{LFI:white:noise:sensitivity:70GHz:units}{134.7\muKs}
\setsymbol{LFI:white:noise:sensitivity:44GHz:units}{164.7\muKs}
\setsymbol{LFI:white:noise:sensitivity:30GHz:units}{143.4\muKs}

\setsymbol{LFI:white:noise:sensitivity:70GHz}{134.7}
\setsymbol{LFI:white:noise:sensitivity:44GHz}{164.7}
\setsymbol{LFI:white:noise:sensitivity:30GHz}{143.4}

\setsymbol{LFI:white:noise:sensitivity:LFI18:Rad:M:units}{512.0\muKs}
\setsymbol{LFI:white:noise:sensitivity:LFI19:Rad:M:units}{581.4\muKs}
\setsymbol{LFI:white:noise:sensitivity:LFI20:Rad:M:units}{590.8\muKs}
\setsymbol{LFI:white:noise:sensitivity:LFI21:Rad:M:units}{455.2\muKs}
\setsymbol{LFI:white:noise:sensitivity:LFI22:Rad:M:units}{492.0\muKs}
\setsymbol{LFI:white:noise:sensitivity:LFI23:Rad:M:units}{507.7\muKs}
\setsymbol{LFI:white:noise:sensitivity:LFI24:Rad:M:units}{462.2\muKs}
\setsymbol{LFI:white:noise:sensitivity:LFI25:Rad:M:units}{413.6\muKs}
\setsymbol{LFI:white:noise:sensitivity:LFI26:Rad:M:units}{478.6\muKs}
\setsymbol{LFI:white:noise:sensitivity:LFI27:Rad:M:units}{277.7\muKs}
\setsymbol{LFI:white:noise:sensitivity:LFI28:Rad:M:units}{312.3\muKs}
\setsymbol{LFI:white:noise:sensitivity:LFI18:Rad:S:units}{465.7\muKs}
\setsymbol{LFI:white:noise:sensitivity:LFI19:Rad:S:units}{555.6\muKs}
\setsymbol{LFI:white:noise:sensitivity:LFI20:Rad:S:units}{623.2\muKs}
\setsymbol{LFI:white:noise:sensitivity:LFI21:Rad:S:units}{564.1\muKs}
\setsymbol{LFI:white:noise:sensitivity:LFI22:Rad:S:units}{534.4\muKs}
\setsymbol{LFI:white:noise:sensitivity:LFI23:Rad:S:units}{542.4\muKs}
\setsymbol{LFI:white:noise:sensitivity:LFI24:Rad:S:units}{399.2\muKs}
\setsymbol{LFI:white:noise:sensitivity:LFI25:Rad:S:units}{392.6\muKs}
\setsymbol{LFI:white:noise:sensitivity:LFI26:Rad:S:units}{418.6\muKs}
\setsymbol{LFI:white:noise:sensitivity:LFI27:Rad:S:units}{302.9\muKs}
\setsymbol{LFI:white:noise:sensitivity:LFI28:Rad:S:units}{285.3\muKs}

\setsymbol{LFI:white:noise:sensitivity:LFI18:Rad:M}{512.0}
\setsymbol{LFI:white:noise:sensitivity:LFI19:Rad:M}{581.4}
\setsymbol{LFI:white:noise:sensitivity:LFI20:Rad:M}{590.8}
\setsymbol{LFI:white:noise:sensitivity:LFI21:Rad:M}{455.2}
\setsymbol{LFI:white:noise:sensitivity:LFI22:Rad:M}{492.0}
\setsymbol{LFI:white:noise:sensitivity:LFI23:Rad:M}{507.7}
\setsymbol{LFI:white:noise:sensitivity:LFI24:Rad:M}{462.2}
\setsymbol{LFI:white:noise:sensitivity:LFI25:Rad:M}{413.6}
\setsymbol{LFI:white:noise:sensitivity:LFI26:Rad:M}{478.6}
\setsymbol{LFI:white:noise:sensitivity:LFI27:Rad:M}{277.7}
\setsymbol{LFI:white:noise:sensitivity:LFI28:Rad:M}{312.3}
\setsymbol{LFI:white:noise:sensitivity:LFI18:Rad:S}{465.7}
\setsymbol{LFI:white:noise:sensitivity:LFI19:Rad:S}{555.6}
\setsymbol{LFI:white:noise:sensitivity:LFI20:Rad:S}{623.2}
\setsymbol{LFI:white:noise:sensitivity:LFI21:Rad:S}{564.1}
\setsymbol{LFI:white:noise:sensitivity:LFI22:Rad:S}{534.4}
\setsymbol{LFI:white:noise:sensitivity:LFI23:Rad:S}{542.4}
\setsymbol{LFI:white:noise:sensitivity:LFI24:Rad:S}{399.2}
\setsymbol{LFI:white:noise:sensitivity:LFI25:Rad:S}{392.6}
\setsymbol{LFI:white:noise:sensitivity:LFI26:Rad:S}{418.6}
\setsymbol{LFI:white:noise:sensitivity:LFI27:Rad:S}{302.9}
\setsymbol{LFI:white:noise:sensitivity:LFI28:Rad:S}{285.3}

\setsymbol{LFI:knee:frequency:70GHz:units}{29.5\mHz}
\setsymbol{LFI:knee:frequency:44GHz:units}{56.2\mHz}
\setsymbol{LFI:knee:frequency:30GHz:units}{113.7\mHz}

\setsymbol{LFI:knee:frequency:70GHz}{29.5}
\setsymbol{LFI:knee:frequency:44GHz}{56.2}
\setsymbol{LFI:knee:frequency:30GHz}{113.7}

\setsymbol{LFI:knee:frequency:LFI18:Rad:M:units}{16.3\mHz}
\setsymbol{LFI:knee:frequency:LFI19:Rad:M:units}{15.1\mHz}
\setsymbol{LFI:knee:frequency:LFI20:Rad:M:units}{18.7\mHz}
\setsymbol{LFI:knee:frequency:LFI21:Rad:M:units}{37.2\mHz}
\setsymbol{LFI:knee:frequency:LFI22:Rad:M:units}{12.7\mHz}
\setsymbol{LFI:knee:frequency:LFI23:Rad:M:units}{34.6\mHz}
\setsymbol{LFI:knee:frequency:LFI24:Rad:M:units}{46.2\mHz}
\setsymbol{LFI:knee:frequency:LFI25:Rad:M:units}{24.9\mHz}
\setsymbol{LFI:knee:frequency:LFI26:Rad:M:units}{67.6\mHz}
\setsymbol{LFI:knee:frequency:LFI27:Rad:M:units}{187.4\mHz}
\setsymbol{LFI:knee:frequency:LFI28:Rad:M:units}{122.2\mHz}
\setsymbol{LFI:knee:frequency:LFI18:Rad:S:units}{17.7\mHz}
\setsymbol{LFI:knee:frequency:LFI19:Rad:S:units}{22.0\mHz}
\setsymbol{LFI:knee:frequency:LFI20:Rad:S:units}{8.7\mHz}
\setsymbol{LFI:knee:frequency:LFI21:Rad:S:units}{25.9\mHz}
\setsymbol{LFI:knee:frequency:LFI22:Rad:S:units}{15.8\mHz}
\setsymbol{LFI:knee:frequency:LFI23:Rad:S:units}{129.8\mHz}
\setsymbol{LFI:knee:frequency:LFI24:Rad:S:units}{100.9\mHz}
\setsymbol{LFI:knee:frequency:LFI25:Rad:S:units}{38.9\mHz}
\setsymbol{LFI:knee:frequency:LFI26:Rad:S:units}{58.9\mHz}
\setsymbol{LFI:knee:frequency:LFI27:Rad:S:units}{104.4\mHz}
\setsymbol{LFI:knee:frequency:LFI28:Rad:S:units}{40.7\mHz}

\setsymbol{LFI:knee:frequency:LFI18:Rad:M}{16.3}
\setsymbol{LFI:knee:frequency:LFI19:Rad:M}{15.1}
\setsymbol{LFI:knee:frequency:LFI20:Rad:M}{18.7}
\setsymbol{LFI:knee:frequency:LFI21:Rad:M}{37.2}
\setsymbol{LFI:knee:frequency:LFI22:Rad:M}{12.7}
\setsymbol{LFI:knee:frequency:LFI23:Rad:M}{34.6}
\setsymbol{LFI:knee:frequency:LFI24:Rad:M}{46.2}
\setsymbol{LFI:knee:frequency:LFI25:Rad:M}{24.9}
\setsymbol{LFI:knee:frequency:LFI26:Rad:M}{67.6}
\setsymbol{LFI:knee:frequency:LFI27:Rad:M}{187.4}
\setsymbol{LFI:knee:frequency:LFI28:Rad:M}{122.2}
\setsymbol{LFI:knee:frequency:LFI18:Rad:S}{17.7}
\setsymbol{LFI:knee:frequency:LFI19:Rad:S}{22.0}
\setsymbol{LFI:knee:frequency:LFI20:Rad:S}{8.7}
\setsymbol{LFI:knee:frequency:LFI21:Rad:S}{25.9}
\setsymbol{LFI:knee:frequency:LFI22:Rad:S}{15.8}
\setsymbol{LFI:knee:frequency:LFI23:Rad:S}{129.8}
\setsymbol{LFI:knee:frequency:LFI24:Rad:S}{100.9}
\setsymbol{LFI:knee:frequency:LFI25:Rad:S}{38.9}
\setsymbol{LFI:knee:frequency:LFI26:Rad:S}{58.9}
\setsymbol{LFI:knee:frequency:LFI27:Rad:S}{104.4}
\setsymbol{LFI:knee:frequency:LFI28:Rad:S}{40.7}

\setsymbol{LFI:slope:70GHz:units}{$-1.03$\mHz}
\setsymbol{LFI:slope:44GHz:units}{$-0.89$\mHz}
\setsymbol{LFI:slope:30GHz:units}{$-0.87$\mHz}

\setsymbol{LFI:slope:70GHz}{$-1.03$}
\setsymbol{LFI:slope:44GHz}{$-0.89$}
\setsymbol{LFI:slope:30GHz}{$-0.87$}

\setsymbol{LFI:slope:LFI18:Rad:M:units}{$-1.04$\mHz}
\setsymbol{LFI:slope:LFI19:Rad:M:units}{$-1.09$\mHz}
\setsymbol{LFI:slope:LFI20:Rad:M:units}{$-0.69$\mHz}
\setsymbol{LFI:slope:LFI21:Rad:M:units}{$-1.56$\mHz}
\setsymbol{LFI:slope:LFI22:Rad:M:units}{$-1.01$\mHz}
\setsymbol{LFI:slope:LFI23:Rad:M:units}{$-0.96$\mHz}
\setsymbol{LFI:slope:LFI24:Rad:M:units}{$-0.83$\mHz}
\setsymbol{LFI:slope:LFI25:Rad:M:units}{$-0.91$\mHz}
\setsymbol{LFI:slope:LFI26:Rad:M:units}{$-0.95$\mHz}
\setsymbol{LFI:slope:LFI27:Rad:M:units}{$-0.87$\mHz}
\setsymbol{LFI:slope:LFI28:Rad:M:units}{$-0.88$\mHz}
\setsymbol{LFI:slope:LFI18:Rad:S:units}{$-1.15$\mHz}
\setsymbol{LFI:slope:LFI19:Rad:S:units}{$-1.00$\mHz}
\setsymbol{LFI:slope:LFI20:Rad:S:units}{$-0.95$\mHz}
\setsymbol{LFI:slope:LFI21:Rad:S:units}{$-0.92$\mHz}
\setsymbol{LFI:slope:LFI22:Rad:S:units}{$-1.01$\mHz}
\setsymbol{LFI:slope:LFI23:Rad:S:units}{$-0.95$\mHz}
\setsymbol{LFI:slope:LFI24:Rad:S:units}{$-0.73$\mHz}
\setsymbol{LFI:slope:LFI25:Rad:S:units}{$-1.16$\mHz}
\setsymbol{LFI:slope:LFI26:Rad:S:units}{$-0.79$\mHz}
\setsymbol{LFI:slope:LFI27:Rad:S:units}{$-0.82$\mHz}
\setsymbol{LFI:slope:LFI28:Rad:S:units}{$-0.91$\mHz}

\setsymbol{LFI:slope:LFI18:Rad:M}{$-1.04$}
\setsymbol{LFI:slope:LFI19:Rad:M}{$-1.09$}
\setsymbol{LFI:slope:LFI20:Rad:M}{$-0.69$}
\setsymbol{LFI:slope:LFI21:Rad:M}{$-1.56$}
\setsymbol{LFI:slope:LFI22:Rad:M}{$-1.01$}
\setsymbol{LFI:slope:LFI23:Rad:M}{$-0.96$}
\setsymbol{LFI:slope:LFI24:Rad:M}{$-0.83$}
\setsymbol{LFI:slope:LFI25:Rad:M}{$-0.91$}
\setsymbol{LFI:slope:LFI26:Rad:M}{$-0.95$}
\setsymbol{LFI:slope:LFI27:Rad:M}{$-0.87$}
\setsymbol{LFI:slope:LFI28:Rad:M}{$-0.88$}
\setsymbol{LFI:slope:LFI18:Rad:S}{$-1.15$}
\setsymbol{LFI:slope:LFI19:Rad:S}{$-1.00$}
\setsymbol{LFI:slope:LFI20:Rad:S}{$-0.95$}
\setsymbol{LFI:slope:LFI21:Rad:S}{$-0.92$}
\setsymbol{LFI:slope:LFI22:Rad:S}{$-1.01$}
\setsymbol{LFI:slope:LFI23:Rad:S}{$-0.95$}
\setsymbol{LFI:slope:LFI24:Rad:S}{$-0.73$}
\setsymbol{LFI:slope:LFI25:Rad:S}{$-1.16$}
\setsymbol{LFI:slope:LFI26:Rad:S}{$-0.79$}
\setsymbol{LFI:slope:LFI27:Rad:S}{$-0.82$}
\setsymbol{LFI:slope:LFI28:Rad:S}{$-0.91$}

\setsymbol{LFI:FWHM:70GHz:units}{13\parcm01}
\setsymbol{LFI:FWHM:44GHz:units}{27\parcm92}
\setsymbol{LFI:FWHM:30GHz:units}{32\parcm65}

\setsymbol{LFI:FWHM:70GHz}{13.01}
\setsymbol{LFI:FWHM:44GHz}{27.92}
\setsymbol{LFI:FWHM:30GHz}{32.65}

\setsymbol{LFI:FWHM:LFI18:units}{13\parcm39}
\setsymbol{LFI:FWHM:LFI19:units}{13\parcm01}
\setsymbol{LFI:FWHM:LFI20:units}{12\parcm75}
\setsymbol{LFI:FWHM:LFI21:units}{12\parcm74}
\setsymbol{LFI:FWHM:LFI22:units}{12\parcm87}
\setsymbol{LFI:FWHM:LFI23:units}{13\parcm27}
\setsymbol{LFI:FWHM:LFI24:units}{22\parcm98}
\setsymbol{LFI:FWHM:LFI25:units}{30\parcm46}
\setsymbol{LFI:FWHM:LFI26:units}{30\parcm31}
\setsymbol{LFI:FWHM:LFI27:units}{32\parcm65}
\setsymbol{LFI:FWHM:LFI28:units}{32\parcm66}

\setsymbol{LFI:FWHM:LFI18}{13.39}
\setsymbol{LFI:FWHM:LFI19}{13.01}
\setsymbol{LFI:FWHM:LFI20}{12.75}
\setsymbol{LFI:FWHM:LFI21}{12.74}
\setsymbol{LFI:FWHM:LFI22}{12.87}
\setsymbol{LFI:FWHM:LFI23}{13.27}
\setsymbol{LFI:FWHM:LFI24}{22.98}
\setsymbol{LFI:FWHM:LFI25}{30.46}
\setsymbol{LFI:FWHM:LFI26}{30.31}
\setsymbol{LFI:FWHM:LFI27}{32.65}
\setsymbol{LFI:FWHM:LFI28}{32.66}

\setsymbol{LFI:FWHM:uncertainty:LFI18:units}{0.170\arcm}
\setsymbol{LFI:FWHM:uncertainty:LFI19:units}{0.174\arcm}
\setsymbol{LFI:FWHM:uncertainty:LFI20:units}{0.170\arcm}
\setsymbol{LFI:FWHM:uncertainty:LFI21:units}{0.156\arcm}
\setsymbol{LFI:FWHM:uncertainty:LFI22:units}{0.164\arcm}
\setsymbol{LFI:FWHM:uncertainty:LFI23:units}{0.171\arcm}
\setsymbol{LFI:FWHM:uncertainty:LFI24:units}{0.652\arcm}
\setsymbol{LFI:FWHM:uncertainty:LFI25:units}{1.075\arcm}
\setsymbol{LFI:FWHM:uncertainty:LFI26:units}{1.131\arcm}
\setsymbol{LFI:FWHM:uncertainty:LFI27:units}{1.266\arcm}
\setsymbol{LFI:FWHM:uncertainty:LFI28:units}{1.287\arcm}

\setsymbol{LFI:FWHM:uncertainty:LFI18}{0.170}
\setsymbol{LFI:FWHM:uncertainty:LFI19}{0.174}
\setsymbol{LFI:FWHM:uncertainty:LFI20}{0.170}
\setsymbol{LFI:FWHM:uncertainty:LFI21}{0.156}
\setsymbol{LFI:FWHM:uncertainty:LFI22}{0.164}
\setsymbol{LFI:FWHM:uncertainty:LFI23}{0.171}
\setsymbol{LFI:FWHM:uncertainty:LFI24}{0.652}
\setsymbol{LFI:FWHM:uncertainty:LFI25}{1.075}
\setsymbol{LFI:FWHM:uncertainty:LFI26}{1.131}
\setsymbol{LFI:FWHM:uncertainty:LFI27}{1.266}
\setsymbol{LFI:FWHM:uncertainty:LFI28}{1.287}

\setsymbol{HFI:center:frequency:100GHz:units}{100\,GHz}
\setsymbol{HFI:center:frequency:143GHz:units}{143\,GHz}
\setsymbol{HFI:center:frequency:217GHz:units}{217\,GHz}
\setsymbol{HFI:center:frequency:353GHz:units}{353\,GHz}
\setsymbol{HFI:center:frequency:545GHz:units}{545\,GHz}
\setsymbol{HFI:center:frequency:857GHz:units}{857\,GHz}

\setsymbol{HFI:center:frequency:100GHz}{100}
\setsymbol{HFI:center:frequency:143GHz}{143}
\setsymbol{HFI:center:frequency:217GHz}{217}
\setsymbol{HFI:center:frequency:353GHz}{353}
\setsymbol{HFI:center:frequency:545GHz}{545}
\setsymbol{HFI:center:frequency:857GHz}{857}

\setsymbol{HFI:Ndetectors:100GHz}{8}
\setsymbol{HFI:Ndetectors:143GHz}{11}
\setsymbol{HFI:Ndetectors:217GHz}{12}
\setsymbol{HFI:Ndetectors:353GHz}{12}
\setsymbol{HFI:Ndetectors:545GHz}{3}
\setsymbol{HFI:Ndetectors:857GHz}{4}

\setsymbol{HFI:FWHM:Maps:100GHz:units}{9\parcm88}
\setsymbol{HFI:FWHM:Maps:143GHz:units}{7\parcm18}
\setsymbol{HFI:FWHM:Maps:217GHz:units}{4\parcm87}
\setsymbol{HFI:FWHM:Maps:353GHz:units}{4\parcm65}
\setsymbol{HFI:FWHM:Maps:545GHz:units}{4\parcm72}
\setsymbol{HFI:FWHM:Maps:857GHz:units}{4\parcm39}
\setsymbol{HFI:FWHM:Maps:100GHz}{9.88}
\setsymbol{HFI:FWHM:Maps:143GHz}{7.18}
\setsymbol{HFI:FWHM:Maps:217GHz}{4.87}
\setsymbol{HFI:FWHM:Maps:353GHz}{4.65}
\setsymbol{HFI:FWHM:Maps:545GHz}{4.72}
\setsymbol{HFI:FWHM:Maps:857GHz}{4.39}

\setsymbol{HFI:beam:ellipticity:Maps:100GHz}{1.15}
\setsymbol{HFI:beam:ellipticity:Maps:143GHz}{1.01}
\setsymbol{HFI:beam:ellipticity:Maps:217GHz}{1.06}
\setsymbol{HFI:beam:ellipticity:Maps:353GHz}{1.05}
\setsymbol{HFI:beam:ellipticity:Maps:545GHz}{1.14}
\setsymbol{HFI:beam:ellipticity:Maps:857GHz}{1.19}

\setsymbol{HFI:FWHM:Mars:100GHz:units}{9\parcm37}
\setsymbol{HFI:FWHM:Mars:143GHz:units}{7\parcm04}
\setsymbol{HFI:FWHM:Mars:217GHz:units}{4\parcm68}
\setsymbol{HFI:FWHM:Mars:353GHz:units}{4\parcm43}
\setsymbol{HFI:FWHM:Mars:545GHz:units}{3\parcm80}
\setsymbol{HFI:FWHM:Mars:857GHz:units}{3\parcm67}

\setsymbol{HFI:FWHM:Mars:100GHz}{9.37}
\setsymbol{HFI:FWHM:Mars:143GHz}{7.04}
\setsymbol{HFI:FWHM:Mars:217GHz}{4.68}
\setsymbol{HFI:FWHM:Mars:353GHz}{4.43}
\setsymbol{HFI:FWHM:Mars:545GHz}{3.80}
\setsymbol{HFI:FWHM:Mars:857GHz}{3.67}

\setsymbol{HFI:beam:ellipticity:Mars:100GHz}{1.18}
\setsymbol{HFI:beam:ellipticity:Mars:143GHz}{1.03}
\setsymbol{HFI:beam:ellipticity:Mars:217GHz}{1.14}
\setsymbol{HFI:beam:ellipticity:Mars:353GHz}{1.09}
\setsymbol{HFI:beam:ellipticity:Mars:545GHz}{1.25}
\setsymbol{HFI:beam:ellipticity:Mars:857GHz}{1.03}

\setsymbol{HFI:CMB:relative:calibration:100GHz}{$\lsim 1\%$}
\setsymbol{HFI:CMB:relative:calibration:143GHz}{$\lsim 1\%$}
\setsymbol{HFI:CMB:relative:calibration:217GHz}{$\lsim 1\%$}
\setsymbol{HFI:CMB:relative:calibration:353GHz}{$\lsim 1\%$}
\setsymbol{HFI:CMB:relative:calibration:545GHz}{}
\setsymbol{HFI:CMB:relative:calibration:857GHz}{}

\setsymbol{HFI:CMB:absolute:calibration:100GHz}{$\lsim 2\%$}
\setsymbol{HFI:CMB:absolute:calibration:143GHz}{$\lsim 2\%$}
\setsymbol{HFI:CMB:absolute:calibration:217GHz}{$\lsim 2\%$}
\setsymbol{HFI:CMB:absolute:calibration:353GHz}{$\lsim 2\%$}
\setsymbol{HFI:CMB:absolute:calibration:545GHz}{}
\setsymbol{HFI:CMB:absolute:calibration:857GHz}{}

\setsymbol{HFI:FIRAS:gain:calibration:accuracy:statistical:100GHz}{}
\setsymbol{HFI:FIRAS:gain:calibration:accuracy:statistical:143GHz}{}
\setsymbol{HFI:FIRAS:gain:calibration:accuracy:statistical:217GHz}{}
\setsymbol{HFI:FIRAS:gain:calibration:accuracy:statistical:353GHz}{2.5\%}
\setsymbol{HFI:FIRAS:gain:calibration:accuracy:statistical:545GHz}{1\%}
\setsymbol{HFI:FIRAS:gain:calibration:accuracy:statistical:857GHz}{0.5\%}

\setsymbol{HFI:FIRAS:gain:calibration:accuracy:systematic:100GHz}{}
\setsymbol{HFI:FIRAS:gain:calibration:accuracy:systematic:143GHz}{}
\setsymbol{HFI:FIRAS:gain:calibration:accuracy:systematic:217GHz}{}
\setsymbol{HFI:FIRAS:gain:calibration:accuracy:systematic:353GHz}{}
\setsymbol{HFI:FIRAS:gain:calibration:accuracy:systematic:545GHz}{7\%}
\setsymbol{HFI:FIRAS:gain:calibration:accuracy:systematic:857GHz}{7\%}

\setsymbol{HFI:FIRAS:zero:point:accuracy:100GHz:units}{0.8\MJysr}
\setsymbol{HFI:FIRAS:zero:point:accuracy:143GHz:units}{}
\setsymbol{HFI:FIRAS:zero:point:accuracy:217GHz:units}{}
\setsymbol{HFI:FIRAS:zero:point:accuracy:353GHz:units}{1.4\MJysr}
\setsymbol{HFI:FIRAS:zero:point:accuracy:545GHz:units}{2.2\MJysr}
\setsymbol{HFI:FIRAS:zero:point:accuracy:857GHz:units}{1.7\MJysr}

\setsymbol{HFI:FIRAS:zero:point:accuracy:100GHz}{0.8}
\setsymbol{HFI:FIRAS:zero:point:accuracy:143GHz}{}
\setsymbol{HFI:FIRAS:zero:point:accuracy:217GHz}{}
\setsymbol{HFI:FIRAS:zero:point:accuracy:353GHz}{1.4}
\setsymbol{HFI:FIRAS:zero:point:accuracy:545GHz}{2.2}
\setsymbol{HFI:FIRAS:zero:point:accuracy:857GHz}{1.7}

\setsymbol{HFI:unit:conversion:100GHz:units}{0.2415\MJysrmK}
\setsymbol{HFI:unit:conversion:143GHz:units}{0.3694\MJysrmK}
\setsymbol{HFI:unit:conversion:217GHz:units}{0.4811\MJysrmK}
\setsymbol{HFI:unit:conversion:353GHz:units}{0.2883\MJysrmK}
\setsymbol{HFI:unit:conversion:545GHz:units}{0.05826\MJysrmK}
\setsymbol{HFI:unit:conversion:857GHz:units}{0.002238\MJysrmK}

\setsymbol{HFI:unit:conversion:100GHz}{0.2415}
\setsymbol{HFI:unit:conversion:143GHz}{0.3694}
\setsymbol{HFI:unit:conversion:217GHz}{0.4811}
\setsymbol{HFI:unit:conversion:353GHz}{0.2883}
\setsymbol{HFI:unit:conversion:545GHz}{0.05826}
\setsymbol{HFI:unit:conversion:857GHz}{0.002238}

\setsymbol{HFI:colour:correction:alpha=-2:V1.01:100GHz}{0.9893}
\setsymbol{HFI:colour:correction:alpha=-2:V1.01:143GHz}{0.9759}
\setsymbol{HFI:colour:correction:alpha=-2:V1.01:217GHz}{1.0007}
\setsymbol{HFI:colour:correction:alpha=-2:V1.01:353GHz}{1.0028}
\setsymbol{HFI:colour:correction:alpha=-2:V1.01:545GHz}{1.0019}
\setsymbol{HFI:colour:correction:alpha=-2:V1.01:857GHz}{0.9889}

\setsymbol{HFI:colour:correction:alpha=0:V1.01:100GHz}{1.0008}
\setsymbol{HFI:colour:correction:alpha=0:V1.01:143GHz}{1.0148}
\setsymbol{HFI:colour:correction:alpha=0:V1.01:217GHz}{0.9909}
\setsymbol{HFI:colour:correction:alpha=0:V1.01:353GHz}{0.9888}
\setsymbol{HFI:colour:correction:alpha=0:V1.01:545GHz}{0.9878}
\setsymbol{HFI:colour:correction:alpha=0:V1.01:857GHz}{1.0014}

\newcommand\papertitle{The millimetre  emission from Planetary Nebulae}

\newcommand{\planck}{\Planck} 

\newcommand{\Spitzer}{\textit{Spitzer\/}}
\newcommand{\IRAS}{\textit{IRAS\/}}

\newcommand{\ISO}{\textit{ISO\/}}
\newcommand{\AKARI}{\textit{Akari\/}}
\newcommand{\MSX}{\textit{MSX\/}}
\newcommand{\WISE}{\textit{WISE\/}}
\newcommand{\WMAP}{\textit{WMAP\/}}
\newcommand{\GALEX}{\textit{GALEX\/}}

\newcommand{\um}{\,$\upmu$m}
\newcommand{\hone}{\ion{H}{i}}
\newcommand{\htwo}{\ion{H}{ii}}

\newcommand{\s}{\scriptsize}

\begin{document}
\author{\small
Planck Collaboration:
M.~Arnaud\inst{64}
\and
F.~Atrio-Barandela\inst{16}
\and
J.~Aumont\inst{52}
\and
C.~Baccigalupi\inst{74}
\and
A.~J.~Banday\inst{80, 9}
\and
R.~B.~Barreiro\inst{58}
\and
E.~Battaner\inst{81, 82}
\and
K.~Benabed\inst{53, 78}
\and
A.~Benoit-L\'{e}vy\inst{21, 53, 78}
\and
J.-P.~Bernard\inst{80, 9}
% ...
}
\institute{\small
APC, AstroParticule et Cosmologie, Universit\'{e} Paris Diderot, CNRS/IN2P3, CEA/lrfu, Observatoire de Paris, Sorbonne Paris Cit\'{e}, 10, rue Alice Domon et L\'{e}onie Duquet, 75205 Paris Cedex 13, France\goodbreak
\and
Aalto University Mets\"{a}hovi Radio Observatory and Dept of Radio Science and Engineering, P.O. Box 13000, FI-00076 AALTO, Finland\goodbreak
\and
African Institute for Mathematical Sciences, 6-8 Melrose Road, Muizenberg, Cape Town, South Africa\goodbreak
\and
Agenzia Spaziale Italiana Science Data Center, Via del Politecnico snc, 00133, Roma, Italy\goodbreak
\and
Agenzia Spaziale Italiana, Viale Liegi 26, Roma, Italy\goodbreak
\and
Astrophysics Group, Cavendish Laboratory, University of Cambridge, J J Thomson Avenue, Cambridge CB3 0HE, U.K.\goodbreak
\and
Astrophysics \& Cosmology Research Unit, School of Mathematics, Statistics \& Computer Science, University of KwaZulu-Natal, Westville Campus, Private Bag X54001, Durban 4000, South Africa\goodbreak
\and
CITA, University of Toronto, 60 St. George St., Toronto, ON M5S 3H8, Canada\goodbreak
\and
CNRS, IRAP, 9 Av. colonel Roche, BP 44346, F-31028 Toulouse cedex 4, France\goodbreak
\and
California Institute of Technology, Pasadena, California, U.S.A.\goodbreak
\and
Centro de Astrobiolog\'{i}a (INTA-CSIC), 28850 Torrej\'{o}n de Ardoz, Madrid, Spain\goodbreak
\and
Computational Cosmology Center, Lawrence Berkeley National Laboratory, Berkeley, California, U.S.A.\goodbreak
\and
DSM/Irfu/SPP, CEA-Saclay, F-91191 Gif-sur-Yvette Cedex, France\goodbreak
\and
DTU Space, National Space Institute, Technical University of Denmark, Elektrovej 327, DK-2800 Kgs. Lyngby, Denmark\goodbreak
\and
D\'{e}partement de Physique Th\'{e}orique, Universit\'{e} de Gen\`{e}ve, 24, Quai E. Ansermet,1211 Gen\`{e}ve 4, Switzerland\goodbreak
\and
Departamento de F\'{\i}sica Fundamental, Facultad de Ciencias, Universidad de Salamanca, 37008 Salamanca, Spain\goodbreak
\and
Departamento de F\'{\i}sica, Universidad de Oviedo, Avda. Calvo Sotelo s/n, Oviedo, Spain\goodbreak
\and
Department of Astrophysics/IMAPP, Radboud University Nijmegen, P.O. Box 9010, 6500 GL Nijmegen, The Netherlands\goodbreak
\and
Department of Physics \& Astronomy, University of British Columbia, 6224 Agricultural Road, Vancouver, British Columbia, Canada\goodbreak
\and
Department of Physics and Astronomy, Dana and David Dornsife College of Letter, Arts and Sciences, University of Southern California, Los Angeles, CA 90089, U.S.A.\goodbreak
\and
Department of Physics and Astronomy, University College London, London WC1E 6BT, U.K.\goodbreak
\and
Department of Physics, Florida State University, Keen Physics Building, 77 Chieftan Way, Tallahassee, Florida, U.S.A.\goodbreak
\and
Department of Physics, Gustaf H\"{a}llstr\"{o}min katu 2a, University of Helsinki, Helsinki, Finland\goodbreak
\and
Department of Physics, Princeton University, Princeton, New Jersey, U.S.A.\goodbreak
\and
Department of Physics, University of California, Santa Barbara, California, U.S.A.\goodbreak
\and
Department of Physics, University of Illinois at Urbana-Champaign, 1110 West Green Street, Urbana, Illinois, U.S.A.\goodbreak
\and
Dipartimento di Fisica e Astronomia G. Galilei, Universit\`{a} degli Studi di Padova, via Marzolo 8, 35131 Padova, Italy\goodbreak
\and
Dipartimento di Fisica e Scienze della Terra, Universit\`{a} di Ferrara, Via Saragat 1, 44122 Ferrara, Italy\goodbreak
\and
Dipartimento di Fisica, Universit\`{a} La Sapienza, P. le A. Moro 2, Roma, Italy\goodbreak
\and
Dipartimento di Fisica, Universit\`{a} degli Studi di Milano, Via Celoria, 16, Milano, Italy\goodbreak
\and
Dipartimento di Fisica, Universit\`{a} degli Studi di Trieste, via A. Valerio 2, Trieste, Italy\goodbreak
\and
Dipartimento di Fisica, Universit\`{a} di Roma Tor Vergata, Via della Ricerca Scientifica, 1, Roma, Italy\goodbreak
\and
Discovery Center, Niels Bohr Institute, Blegdamsvej 17, Copenhagen, Denmark\goodbreak
\and
Dpto. Astrof\'{i}sica, Universidad de La Laguna (ULL), E-38206 La Laguna, Tenerife, Spain\goodbreak
\and
European Space Agency, ESAC, Planck Science Office, Camino bajo del Castillo, s/n, Urbanizaci\'{o}n Villafranca del Castillo, Villanueva de la Ca\~{n}ada, Madrid, Spain\goodbreak
\and
European Space Agency, ESTEC, Keplerlaan 1, 2201 AZ Noordwijk, The Netherlands\goodbreak
\and
Facolt\`{a} di Ingegneria, Universit\`{a} degli Studi e-Campus, Via Isimbardi 10, Novedrate (CO), 22060, Italy\goodbreak
\and
Harvard-Smithsonian Center for Astrophysics, 60 Garden Street, Cambridge, MA 02138, U.S.A.\goodbreak
\and
Helsinki Institute of Physics, Gustaf H\"{a}llstr\"{o}min katu 2, University of Helsinki, Helsinki, Finland\goodbreak
\and
INAF - Osservatorio Astrofisico di Catania, Via S. Sofia 78, Catania, Italy\goodbreak
\and
INAF - Osservatorio Astronomico di Padova, Vicolo dell'Osservatorio 5, Padova, Italy\goodbreak
\and
INAF - Osservatorio Astronomico di Roma, via di Frascati 33, Monte Porzio Catone, Italy\goodbreak
\and
INAF - Osservatorio Astronomico di Trieste, Via G.B. Tiepolo 11, Trieste, Italy\goodbreak
\and
INAF Istituto di Radioastronomia, Via P. Gobetti 101, 40129 Bologna, Italy\goodbreak
\and
INAF/IASF Bologna, Via Gobetti 101, Bologna, Italy\goodbreak
\and
INAF/IASF Milano, Via E. Bassini 15, Milano, Italy\goodbreak
\and
INFN, Sezione di Bologna, Via Irnerio 46, I-40126, Bologna, Italy\goodbreak
\and
INFN, Sezione di Roma 1, Universit\`{a} di Roma Sapienza, Piazzale Aldo Moro 2, 00185, Roma, Italy\goodbreak
\and
INFN/National Institute for Nuclear Physics, Via Valerio 2, I-34127 Trieste, Italy\goodbreak
\and
Imperial College London, Astrophysics group, Blackett Laboratory, Prince Consort Road, London, SW7 2AZ, U.K.\goodbreak
\and
Infrared Processing and Analysis Center, California Institute of Technology, Pasadena, CA 91125, U.S.A.\goodbreak
\and
Institut d'Astrophysique Spatiale, CNRS (UMR8617) Universit\'{e} Paris-Sud 11, B\^{a}timent 121, Orsay, France\goodbreak
\and
Institut d'Astrophysique de Paris, CNRS (UMR7095), 98 bis Boulevard Arago, F-75014, Paris, France\goodbreak
\and
Institute for Space Sciences, Bucharest-Magurale, Romania\goodbreak
\and
Institute of Astronomy, University of Cambridge, Madingley Road, Cambridge CB3 0HA, U.K.\goodbreak
\and
Institute of Theoretical Astrophysics, University of Oslo, Blindern, Oslo, Norway\goodbreak
\and
Instituto de Astrof\'{\i}sica de Canarias, C/V\'{\i}a L\'{a}ctea s/n, La Laguna, Tenerife, Spain\goodbreak
\and
Instituto de F\'{\i}sica de Cantabria (CSIC-Universidad de Cantabria), Avda. de los Castros s/n, Santander, Spain\goodbreak
\and
Jet Propulsion Laboratory, California Institute of Technology, 4800 Oak Grove Drive, Pasadena, California, U.S.A.\goodbreak
\and
Jodrell Bank Centre for Astrophysics, Alan Turing Building, School of Physics and Astronomy, The University of Manchester, Oxford Road, Manchester, M13 9PL, U.K.\goodbreak
\and
Kavli Institute for Cosmology Cambridge, Madingley Road, Cambridge, CB3 0HA, U.K.\goodbreak
\and
LAL, Universit\'{e} Paris-Sud, CNRS/IN2P3, Orsay, France\goodbreak
\and
LERMA, CNRS, Observatoire de Paris, 61 Avenue de l'Observatoire, Paris, France\goodbreak
\and
Laboratoire AIM, IRFU/Service d'Astrophysique - CEA/DSM - CNRS - Universit\'{e} Paris Diderot, B\^{a}t. 709, CEA-Saclay, F-91191 Gif-sur-Yvette Cedex, France\goodbreak
\and
Laboratoire Traitement et Communication de l'Information, CNRS (UMR 5141) and T\'{e}l\'{e}com ParisTech, 46 rue Barrault F-75634 Paris Cedex 13, France\goodbreak
\and
Laboratoire de Physique Subatomique et de Cosmologie, Universit\'{e} Joseph Fourier Grenoble I, CNRS/IN2P3, Institut National Polytechnique de Grenoble, 53 rue des Martyrs, 38026 Grenoble cedex, France\goodbreak
\and
Laboratoire de Physique Th\'{e}orique, Universit\'{e} Paris-Sud 11 \& CNRS, B\^{a}timent 210, 91405 Orsay, France\goodbreak
\and
Lawrence Berkeley National Laboratory, Berkeley, California, U.S.A.\goodbreak
\and
Max-Planck-Institut f\"{u}r Astrophysik, Karl-Schwarzschild-Str. 1, 85741 Garching, Germany\goodbreak
\and
National University of Ireland, Department of Experimental Physics, Maynooth, Co. Kildare, Ireland\goodbreak
\and
Niels Bohr Institute, Blegdamsvej 17, Copenhagen, Denmark\goodbreak
\and
Observational Cosmology, Mail Stop 367-17, California Institute of Technology, Pasadena, CA, 91125, U.S.A.\goodbreak
\and
Optical Science Laboratory, University College London, Gower Street, London, U.K.\goodbreak
\and
SISSA, Astrophysics Sector, via Bonomea 265, 34136, Trieste, Italy\goodbreak
\and
School of Physics and Astronomy, Cardiff University, Queens Buildings, The Parade, Cardiff, CF24 3AA, U.K.\goodbreak
\and
Space Sciences Laboratory, University of California, Berkeley, California, U.S.A.\goodbreak
\and
Special Astrophysical Observatory, Russian Academy of Sciences, Nizhnij Arkhyz, Zelenchukskiy region, Karachai-Cherkessian Republic, 369167, Russia\goodbreak
\and
UPMC Univ Paris 06, UMR7095, 98 bis Boulevard Arago, F-75014, Paris, France\goodbreak
\and
Universidad de Chile, Casilla 36-D, Santiago, Chile\goodbreak
\and
Universit\'{e} de Toulouse, UPS-OMP, IRAP, F-31028 Toulouse cedex 4, France\goodbreak
\and
University of Granada, Departamento de F\'{\i}sica Te\'{o}rica y del Cosmos, Facultad de Ciencias, Granada, Spain\goodbreak
\and
University of Granada, Instituto Carlos I de F\'{\i}sica Te\'{o}rica y Computacional, Granada, Spain\goodbreak
% ...
}

\title{\Planck\ intermediate results. XVIII \\ The millimetre and sub-millimetre emission from Planetary Nebulae}



\abstract{
Late stages of stellar evolution are characterized by copious mass-loss events whose signature is the formation of circumstellar envelopes (CSE). \Planck\ multi-frequency measurements have provided relevant information on a sample of Galactic planetary nebulae (PNe) in the important and relatively unexplored observational band between 30 and 857\GHz. \Planck\ enables the assembly of comprehensive PNe spectral energy distributions (SEDs) from radio {\bf to} far-infrared frequencies.
Modelling of the derived SEDs provides us with information on physical properties of CSEs and the mass content of both main components: ionised gas, traced by the free-free emission at cm--mm waves; and thermal dust, traced by the millimetre and far-IR emission. In particular, the amount  of ionised gas and dust has been derived here. Such quantities have also been estimated for the very young PN CRL\,618, where the strong variability observed in its radio and millimetre emission has previously prevented the construction of its SED.
A morphological study of the Helix Nebula has also been performed. \Planck\ maps reveal, for the first time, the spatial distribution of the dust inside the envelope, allowing us to identify different components, the most interesting of which is a very extended component (up to 1\,pc) that may be related to a region where the slow expanding envelope is interacting with the surrounding interstellar medium.}

\keywords{Planetary nebulae: general -- Radio continuum: ISM -- Submillimetre: ISM}
\authorrunning{}
\titlerunning{\papertitle}

\maketitle

\section{Introduction} \label{sec:intro}
The final phases of low to intermediate mass stars are characterized by periods of high mass loss that lead to the formation of dense circumstellar envelopes (CSEs), where physical conditions are ideal for dust to condense (during the Asymptotic Giant Branch, or AGB phase). Such envelopes can be very massive and, in some cases, the central object can be completely optically obscured. Eventually the mass loss stops and the central star becomes visible as the dusty shell disperses (this is the proto-planetary nebula, or PPN phase). During the subsequent evolutionary phases, the central star moves toward higher temperatures and, once the stellar temperature is high enough to ionise the surrounding medium, the object becomes a planetary nebula (PN) \citep{Kwok2008}.

PNe are usually surrounded by a dusty envelope that is a remnant of the previous evolutionary phases and is partly ionised by the UV radiation from the central star. The material surrounding the central object consequently has quite a complex distribution, consisting of concentric shells. In these shells, the level of ionisation decreases with the distance to the central star: highly ionised species are close to the star, while the outer part of the nebula is characterized by molecules and dust. This characteristic circumstellar environment implies the presence of two important components in the spectral energy distribution (SED) of a typical PN, whose major contributions fall in the spectral range from the far-IR to the radio region. The ionised fraction of the CSE can be traced by its free-free emission, which makes PNe bright Galactic radio sources, with some of them reaching flux densities  up to a  few Jy. Dust thermally re-radiates the absorbed stellar light, showing a clear signature in the far-infrared (far-IR) spectrum, i.e., an IR excess in the colour-colour diagram created with data from the Infrared Astronomical Satellite (IRAS). Such a contribution is typically of the order of 40\,\% of the total flux from a PN \citep{Zhang1991}, and is larger in young PNe, since in more evolved PNe the circumstellar material has already dispersed.

PNe and their progenitors are considered to be among the major sources of recycled material into the ISM and for this reason, they are regarded as key objects for studying the chemical evolution of the Galaxy. Before being released into the ISM, significant processing of the material contained in the PN envelopes is expected. Gas and dust are exposed to a very harsh environment: shocks will occur when the fast outflows developing at the beginning of the PN phase overtake the slow expanding AGB envelope, and the UV radiation field radiated by the central star could be very intense, with consequences on the ionisation/recombination equilibrium of the gas and on the dusty/molecular content of the envelope \citep{Hora2009}. It is therefore very important to establish not only how much of this processed material is returned to the ISM after the nebula disperses, but also its general properties and dominant chemistry.

The aim of this work is to derive the physical characteristics of a sample of Galactic PNe, taking advantage of the unique frequency coverage provided by the \Planck\footnote{\Planck\ is a project of the European Space Agency -- ESA -- with instruments provided by two scientific Consortia funded by ESA member states (in particular the lead countries: France and Italy) with contributions from NASA (USA), and telescope reflectors provided in a collaboration between ESA and a scientific Consortium led and funded by Denmark.} measurements, which trace both the ionised and the dust components. We model the SEDs with particular attention to the continuum from the mid-IR to the radio. Results from such modelling will be used to derive important parameters of PN envelopes, i.e., the total mass, the ionised fraction and the properties of the dust component. In some cases, hints on the extended morphology of PN envelopes can be derived from a direct inspection of the \Planck\ maps.

As a starting point, we have compiled a master catalogue of Galactic PNe for which pre-existing 30\GHz\ and/or 43\GHz\ measurements are available. Our catalogue, based on single-dish measurements from the Torun (30\GHz, \citealp{Pazderska2009}) and Noto (43\GHz, \citealp{Umana2008a}) surveys, consists of 119 PNe, and covers a large parameter space in terms of location with respect to the Galactic plane, distance, and evolutionary stage. Typical numbers for the surveys are a FWHM for the beam of 72\arcs\ and an rms noise of 5\,mJy for the One Centimetre Receiver Array-prototype (OCRA-p) observations, and a FWHM for the beam of 52\arcs\ and rms noise of 70\,mJy for the Noto telescope survey.

This paper is organized as follows. Observations, consisting of \Planck\ and ancillary data, are presented in Sect. \ref{sec:obs}, while the methods to extract fluxes and results are described in Sect. \ref{sec:sed}. The adopted SED modelling and derived physical properties of the detected targets are illustrated in Sect. \ref{sec:properties}, while Sects. \ref{sec:crl618} and \ref{sec:helix} focus on two targets, namely CLR\,618 and NGC\,7293 (the Helix Nebula, Helix hereafter), whose characterization appear to be particularly interesting. Section \ref{sec:conclusions} concludes.

%
%
\begin{table*}
\begingroup
\newdimen\tblskip \tblskip=5pt
\caption{Non-colour-corrected flux densities (Jy) from Planck Catalog of Compact Sources \citep[PCCS;][]{planck2013-p05}.
}
\label{tab:fluxes}
\nointerlineskip
\vskip -3mm
\footnotesize
\setbox\tablebox=\vbox{
 \newdimen\digitwidth 
 \setbox0=\hbox{\rm 0} 
 \digitwidth=\wd0 
 \catcode`*=\active 
 \def*{\kern\digitwidth}
 \newdimen\signwidth 
 \setbox0=\hbox{+} 
 \signwidth=\wd0 
 \catcode`!=\active 
 \def!{\kern\signwidth}
 \halign{\hbox to 1in{#\leaderfil}\tabskip .35em&
 \hfil#\hfil&
 \hfil#\hfil&
 \hfil#\hfil& 
 \hfil#\hfil&
 \hfil#\hfil&
 \hfil#\hfil&
 \hfil#\hfil&
 \hfil#\hfil&
 \hfil#\hfil&
 \hfil#\hfil&
#\hfil\tabskip 0pt\cr
 \noalign{\doubleline\vskip 2pt}
 \omit Source \hfil& Coordinates& \multispan9\hfil Frequency (GHz) \hfil& Ancillary Data\cr
\noalign{\vskip 4pt\hrule\vskip 6pt}
 \omit& Galactic& 28.4& 44.1& 70.4& 100& 143& 217& 353& 545& 857& Refs.\cr
\noalign{\vskip 4pt\hrule\vskip 6pt}
\omit NGC\,6369$^{\phantom \dag}_{\phantom \dag}$& \s002.432+05.847& \s1.9\phantom{0}$\pm$0.2\phantom{0}& \s1.7$\pm$0.3\phantom{0}& \s1.27$\pm$0.15& \s1.5\phantom{0}$\pm$0.1\phantom{0}& \s1.27$\pm$0.06& \s1.35$\pm$0.09& \s\phantom{0}2.4\phantom{0}$\pm$0.3\phantom{0}& \s\phantom{0}7.0\phantom{0}$\pm$0.9\phantom{0}& \dots& \s$^{3,~8,~12,~15,~16,~17,~19,~21,~22,~29,~30,~34,~40,~45,~47}_{50,~54}$	\cr
\omit NGC\,6572$^{\phantom \dag}_{\phantom \dag}$& \s034.623+11.848& \s1.1\phantom{0}$\pm$0.1\phantom{0}& \dots& \s1.2\phantom{0}$\pm$0.2\phantom{0}& \s0.96$\pm$0.06& \s0.92$\pm$0.04& \s0.81$\pm$0.04& \s0.81$\pm$0.09& \dots& \dots& \s{$^{1,~2,~3,~4,~5,~6,~7,~8,~9,~10,~13,~16,~17,~18,~19,~21,~22,~23}_{26,~29,~30,~32,~33,~34,~35,~38,~39,~48,~50,~54,~55,~56}$}\cr
\omit NGC\,7293$^{\mathrm a}$ $^{\phantom \dag}_{\phantom \dag}$& \s036.161$-$57.118& \s0.8\phantom{0}$\pm$0.1\phantom{0}& \s1.0$\pm$0.2\phantom{0}& \s0.88$\pm$0.15& \s1.6\phantom{0}$\pm$0.5\phantom{0}& \s1.6\phantom{0}$\pm$0.5\phantom{0}& \s3.4\phantom{0}$\pm$0.6\phantom{0}& \s10\phantom{.00}$\pm$1\phantom{.00}& \s26\phantom{.00}$\pm$3\phantom{.00}& \s73\phantom{.0}$\pm$9\phantom{.0}& \s{$^{2,~3,~6,~7,~16,~17,~18,~19,~23,~29,~52}$}\cr
\omit NGC\,7009$^{\phantom \dag}_{\phantom \dag}$& \s037.762$-$34.571& \s0.6\phantom{0}$\pm$0.1\phantom{0}& \dots& \s0.8\phantom{0}$\pm$0.2\phantom{0}& \s0.54$\pm$0.06& \s0.40$\pm$0.04& \s0.42$\pm$0.04& \dots& \dots& \dots& \s{$^{2,~3,~6,~8,~9,~16,~17,~21,~26,~29,~30,~41,~45,~47,~50,~51}_{54,~55,~57,~59}$}\cr
\omit NGC\,6853$^{\phantom \dag}_{\phantom \dag}$& \s060.836$-$03.696& \dots& \dots& \s0.97$\pm$0.15& \s0.87$\pm$0.07& \s0.65$\pm$0.05& \dots& \dots& \dots& \dots& \s{$^{1,~2,~3,~4,~5,~6,~7,~8,~9,~10,~13,~16,~17,~18,~20,~23,~38}_{50}$}\cr
\omit NGC\,6720$^{\phantom \dag}_{\phantom \dag}$& \s063.170+13.978& \dots& \dots& \dots& \dots& \s0.31$\pm$0.04& \s0.33$\pm$0.03& \s\phantom{0}0.64$\pm$0.06& \s\phantom{0}1.6\phantom{0}$\pm$0.1\phantom{0}& \s\phantom{0}4.7$\pm$0.2& \s{$^{1,~2,~5,~6,~7,~8,~9,~10,~14,~16,~18,~20,~21,~32,~33,~38,~39}_{48,~50,~55,~60}$}\cr
\omit NGC\,6826$^{\phantom \dag}_{\phantom \dag}$& \s083.568+12.792& \s0.35$\pm$0.08& \dots& \dots& \s0.32$\pm$0.04& \s0.25$\pm$0.03& \s0.26$\pm$0.02& \dots& \dots& \dots& \s{$^{6,~8,~10,~12,~16,~17,~21,~33,~38,~39,~48,~50,~55,~60}$}\cr
\omit NGC\,7027$^{\phantom \dag}_{\phantom \dag}$& \s084.930$-$03.496& \s5.0\phantom{0}$\pm$0.9\phantom{0}& \s4.7$\pm$0.6\phantom{0}& \s4.9\phantom{0}$\pm$0.2\phantom{0}& \s4.4\phantom{0}$\pm$0.1\phantom{0}& \s4.25$\pm$0.06& \s4.3\phantom{0}$\pm$0.1\phantom{0}& \s\phantom{0}5.4\phantom{0}$\pm$0.3\phantom{0}& \s\phantom{0}6.3\phantom{0}$\pm$0.7\phantom{0}& \dots& \s{$^{13,~24,~25,~27,~28,~31,~36,~42,~43,~44,~46,~56}$}\cr
\omit NGC\,6543$^{\phantom \dag}_{\phantom \dag}$& \s096.468+29.954& \s0.74$\pm$0.08& \s0.9$\pm$0.15& \s0.67$\pm$0.09& \s0.63$\pm$0.04& \s0.52$\pm$0.02& \s0.54$\pm$0.02& \s\phantom{0}0.59$\pm$0.04& \s\phantom{0}0.84$\pm$0.07& \s\phantom{0}2.6$\pm$0.2& \s{$^{2,~4,~6,~7,~8,~16,~17,~21,~22,~27,~38,~39,~47,~48,~49,~50}_{53,~55,~60}$}\cr
\omit NGC\,40$^{\phantom \dag}_{\phantom \dag}$& \s120.016+09.868& \dots& \dots& \s0.78$\pm$0.15& \s0.45$\pm$0.06& \s0.36$\pm$0.04& \s0.30$\pm$0.04& \dots& \dots& \dots& \s{$^{6,~10,~12,~16,~17,~38,~39,~48,~50,~60}$}\cr
\omit CRL\,618$^{\phantom \dag}_{\phantom \dag}$& \s166.446$-$06.527& \s0.7\phantom{0}$\pm$0.2\phantom{0}& \s1.4$\pm$0.2\phantom{0}& \s1.9\phantom{0}$\pm$0.2\phantom{0}& \s2.40$\pm$0.08& \s2.67$\pm$0.05& \s3.00$\pm$0.07& \s\phantom{0}4.9\phantom{0}$\pm$0.2\phantom{0}& \s10.4$\pm$0.3\phantom{0}& \s24.3$\pm$0.8& \s{$^{32,~56}$}\cr
\omit IC\,418$^{\phantom \dag}_{\phantom \dag}$& \s215.212$-$24.283& \s1.6\phantom{0}$\pm$0.1\phantom{0}& \s1.5$\pm$0.2\phantom{0}& \s1.4\phantom{0}$\pm$0.2\phantom{0}& \s1.18$\pm$0.07& \s1.02$\pm$0.04& \s0.94$\pm$0.03& \s\phantom{0}0.84$\pm$0.08& \dots& \dots& \s{$^{2,~3,~4,~6,~7,~8,~9,~10,~11,~13,~15,~16,~17,~19,~21,~22,~30}_{35,~37,~42,~45,~50,~54,~55,~56,~58,~59}$}\cr
\omit NGC\,3242$^{\phantom \dag}_{\phantom \dag}$& \s261.051+32.050& \s0.6\phantom{0}$\pm$0.1\phantom{0}& \dots& \dots& \s0.51$\pm$0.06& \s0.48$\pm$0.04& \s0.37$\pm$0.03& \s\phantom{0}0.44$\pm$0.06& \dots& \dots& \s{$^{2,~3,~4,~6,~8,~9,~10,~11,~12,~13,~16,~17,~21,~22,~29,~35}_{37,~45,~50,~51,~54,~55,~56,~59}$}\cr
\noalign{\vskip 3pt\hrule\vskip 4pt}
}}
\endPlancktablewide
\vspace{-2mm}
\begin{list}{}{}
\item[$^{\mathrm a}$] Helix Nebula
\item[References:]
(1) \citealp{davies_etal65};
(2) \citealp{menon_terzian65};
(3) \citealp{slee_orchiston65};
(4) \citealp{khromov_66};
(5) \citealp{terzian66};
(6) \citealp{davies_etal67};
(7) \citealp{hughes_67};
(8) \citealp{thompson_etal67};
(9) \citealp{terzian68};
(10) \citealp{kaftan-kassim_69};
(11) \citealp{le_marne_69};
(12) \citealp{ribes69}; 
(13) \citealp{terzian69}; 
(14) \citealp{colla_etal70};
(15) \citealp{rubin_70};
(16) \citealp{thomasson_davies70};
(17) \citealp{higgs_71};
(18) \citealp{aller_milne72};
(19) \citealp{Higgs1973};
(20) \citealp{terzian_dickey73}; 
(21) \citealp{sistla_etal74};
(22) \citealp{terzian_etal74};
(23) \citealp{milne_aller75};
(24) \citealp{Telesco1977};
(25) \citealp{Elias1978};
(26) \citealp{cohen_barlow1980};
(27) \citealp{Moseley1980};
(28) \citealp{Ulich1981};
(29) \citealp{calabretta82};
(30) \citealp{milne_aller82};
(31) \citealp{Gee1984};
(32) \citealp{turner_terzian84};
(33) \citealp{bennet_etal86};
(34) \citealp{gathier_etal86};
(35) \citealp{taylor_etal87};
(36) \citealp{steppe_etal88}; 
(37) \citealp{wright_otrupcek90};
(38) \citealp{becker_etal91};
(39) \citealp{gregory_condon1991}; 
(40) \citealp{large_etal91};
(41) \citealp{wright_etal91};
(42) \citealp{Hoare1992};
(43) \citealp{knapp_etal93}; 
(44) \citealp{altenhoff_etal94}; 
(45) \citealp{griffith_etal94};
(46) \citealp{sandelll94}; 
(47) \citealp{douglas_etal96};
(48) \citealp{gregory_etal1996}; 
(49) \citealp{Rengelink_etal97};
(50) \citealp{condon_kaplan98};
(51) \citealp{De_Breuck_etal02};
(52) \citealp{Casassus2004}; 
(53) \citealp{Klaas2006};
(54) \citealp{casassus_etal07}; 
(55) \citealp{healey_etal07};
(56) \citealp{DiFrancesco2008};
(57) \citealp{vollmer_etal08};
(58) \citealp{wright_etal09};
(59) \citealp{murphy_etal10};
(60) \citealp{vollmer_etal10}.

\end{list}
\endgroup
\end{table*}

\section{Observations} \label{sec:obs}
\subsection{\planck\ data} \label{sec:planck}

\Planck\ \citep{tauber2010a, planck2011-1.1} is the third generation space mission to measure the anisotropy of the cosmic microwave background (CMB). It observes the sky in nine frequency bands covering 30--857\GHz\ with high sensitivity and angular resolution from 31\arcm\ to 5\arcm. The Low Frequency Instrument LFI; \citep{Mandolesi2010, Bersanelli2010, planck2011-1.4} covers the 30, 44, and 70\GHz\ bands with amplifiers cooled to 20\,\hbox{K}. The High Frequency Instrument (HFI; \citealt{Lamarre2010, planck2011-1.5}) covers the 100, 143, 217, 353, 545, and 857\GHz\ bands with bolometers cooled to 0.1\,\hbox{K}. Polarization is measured in all but the highest two bands \citep{Leahy2010, Rosset2010}. A combination of radiative cooling and three mechanical coolers produces the temperatures needed for the detectors and optics \citep{planck2011-1.3}. Two data processing centres (DPCs) check and calibrate the data and make maps of the sky \citep{planck2011-1.7, planck2011-1.6}. The sensitivity, angular resolution, and frequency coverage of \Planck\ make it a powerful instrument for Galactic and extragalactic astrophysics as well as cosmology. Early astrophysics results are given in 
\citealp{planck2011-5.1a,planck2011-5.1b,planck2011-5.2a,planck2011-5.2b,planck2011-5.2c,planck2011-6.1,planck2011-6.2,planck2011-6.3a,planck2011-6.4a,planck2011-6.4b,planck2011-6.6,planck2011-7.0,planck2011-7.2,planck2011-7.3,planck2011-7.7a,planck2011-7.7b,planck2011-7.12,planck2011-7.13,planck2011-5.1c},
based on data taken between 13~August 2009 and 7~June 2010. Intermediate astrophysics results are now being presented in a series of papers based on data taken between 13~August 2009 and 27~November 2010.

For this work we use the \Planck\ catalogue of compact sources \citep[PCCS;][]{planck2013-p05} and the \Planck\ maps from the 2013 distribution of released products \citep{planck2013-p01}. Both products are based on data acquired during the ``nominal'' operations period from 13 August 2009 to 27 November 2010, and are available from the Planck Legacy Archive\footnote{\url{http://www.sciops.esa.int/index.php?project=planck&page=Planck_Legacy_Archive}}. For  the \Planck\ maps we use here,  the CMB thermodynamic units were converted to Rayleigh-Jeans brightness temperature units using the standard conversion factors described by \citet{planck2013-p01}. Some contamination due to CO emission is expected in the 100, 217, and, at a lower level, 353\GHz\ \Planck\ channels \citep{planck2011-1.7}, although this was not taken into account in the following analysis. However, if some degree of CO contamination is present in our data, it is likely to amount only to a tiny fraction of the measured flux densities. In fact,  as can be seen in the following analysis, in most cases the data points at each frequency are in very good agreement with the model and with the ancillary data, which only treat the continuum emission.

\subsection{Ancillary Data} \label{sec:ancillary}
To complete the SED for each of our selected targets, radio data from the literature (references reported in Table~\ref{tab:fluxes}), as well as data from most infrared and radio all-sky surveys, such as the catalogues from the \WMAP 7-year data release \citep{Gold2011}, \IRAS\ \citep{helou_walker88}, \MSX\ \citep{egan_etal03}, \AKARI\ \citep{murakami_etal07}, and \WISE\ \citep{wright_etal10}, have been collated.

To confirm the reliability of the IR and sub-mm flux densities,
ancillary measurements in these wavelength ranges have been compared with the catalogue measurements for NGC\,7027, a
well-known radio/infrared flux-density calibrator.
We found that the \IRAS\ field around NGC\,7027 is unavailable, but the \MSX\ and \AKARI\ measurements
match into a smooth SED with the  SCUBA \citep{DiFrancesco2008} and the other ancillary targeted observations. 
The use of the \WISE\ data needs instead some caution,
because the sources studied in this paper can be too  bright in the \WISE\  bands,
causing heavy saturation.
Whenever  \WISE\ data are available, we use them after 
checking in each band the flags giving the fraction of pixels affected by saturation, a spurious detection or  possible contamination.

For some sources of our sample, further sub-mm measurements are also available.
If the aperture sizes are reported, as done for example by \citet{Hoare1992}, we used the data only for those sources with angular sizes (given in Table~\ref{tab:ff}) smaller than the aperture. 
Additional information about the mm and sub-mm ancillary data is given in Table~\ref{tab:info}.

The uncertainty estimates taken from the catalogues are of the order of 10\,\% for \IRAS, 5\,\% for \MSX, 10\,\% for \AKARI\ and 3\,\% for \WISE, while for targeted observations we refer to the original papers. For one source, the Helix, we have also used IRIS\footnote{\url{http://irsa.ipac.caltech.edu/data/IRIS/}} maps, which were retrieved via the on-line services of the IPAC Infrared Science Archive (IRSA). These are a new generation of \IRAS\ images, which benefit from a better zodiacal light subtraction, as well as better calibration and destriping. The roughly 4\arcm\ resolution of the IRIS maps matches well with the high-frequency bands of \Planck. Details on the whole processing and characterization of the IRIS data can be found in \citet{Miville05}.

\section{Building the spectral energy distribution (SED)} \label{sec:sed}

Compact sources reported in the PCCS were detected in each frequency channel map using a detection pipeline based on the Mexican Hat Wavelet 2 (MHW2) algorithm \citep{Gonzalez2006, Lopez-Caniego2006}. Two independent implementations of the the MHW2 algorithm have been used by the LFI and HFI DPCs. More details on the PCCS can be found in \citet{planck2013-p05}. In order to build the SEDs of the PNe in our sample, the PCCS was queried with a searching radius of $30^{\prime}$ around each PN position as reported in our master catalogue. We assume a positional coincidence if a \Planck\ source is found within the \Planck\ beam ($\theta_\mathrm{beam}$), where $\theta_\mathrm{beam}$ is function of the channel's frequency.

The PCCS provides multiple estimates of the flux density for each source. In our analysis we use the Detection pipeline photometry (DETFLUX), which assumes that the sources are point-like. As previously described, a typical PN SED consists of two main components, the free-free emission from ionised gas and thermal emission from dust, even though an extra contribution (sometimes called anomalous microwave emission) has been claimed \citep{casassus_etal07}. In the case of the LFI channels we expect the emission from PNe to arise essentially from the ionised gas (free-free emission). The dimensions of the associated radio emission are independently known and the PNe can be quite confidently considered as point sources compared to $\theta_\mathrm{beam}$. In the case of the HFI channels, because of the lack of knowledge of the extension of the dusty CSEs around our sources, and because of the smaller size of $\theta_\mathrm{beam}$ at such frequencies, we perform a posteriori visual inspection of the environment of each detected source in the corresponding \Planck\ map to ensure that the assumption of being point-like holds, and also to evaluate the influence of any confusion from diffuse emission not related to the target source.

\begin{figure}
\begin{center}
\includegraphics[width=88mm]{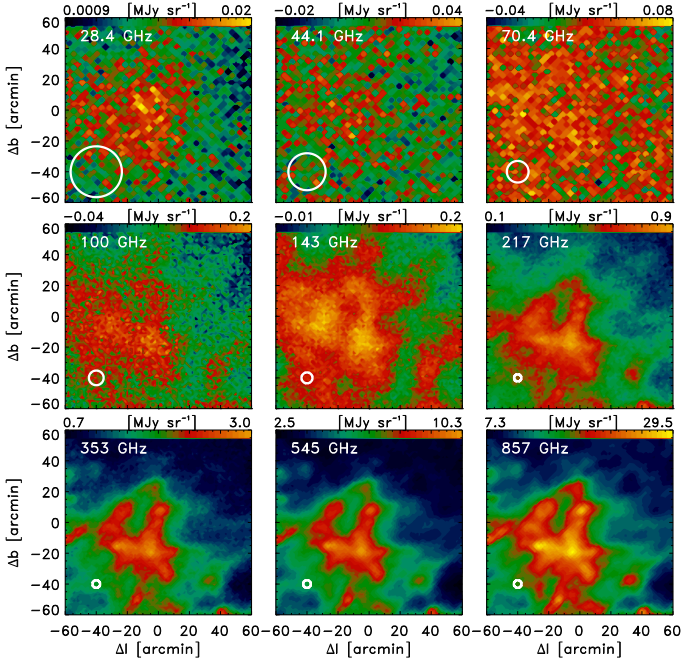}\\
\caption{\Planck\ maps of NGC\,1514. The central PN is embedded in a region of diffuse emission that is not clearly related to the source, which prevents us from obtaining a good photometric measurement. The average FWHM of the effective beam is shown in the bottom-left corner of each map.} 
\label{fig:ngc1514}
\end{center}
\end{figure}

\begin{figure}
\centering
\begin{center}
\includegraphics[width=88mm]{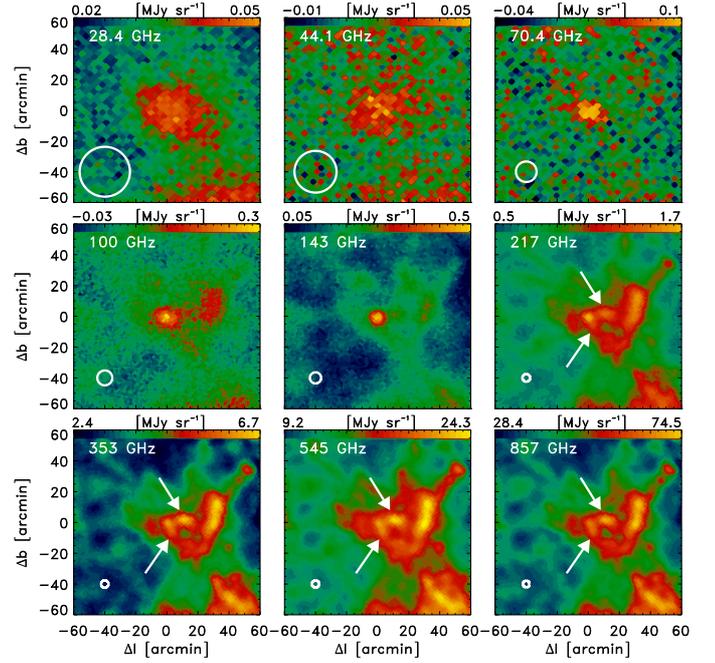}\\
\caption{\Planck\ maps of NGC\,6369. An extended structure, marked by the arrows, appears to surround the PN, mostly visible in the HFI channels. This may constitute a hint for the presence of a dusty halo. However, the central component is well detected in several channels, allowing us to calculate its flux. The average FWHM of the effective beam is shown in the bottom-left corner of each map.}
\label{fig:ngc6369}
\end{center}
\end{figure}

\begin{figure}
\centering
\begin{center}
\includegraphics[width=88mm]{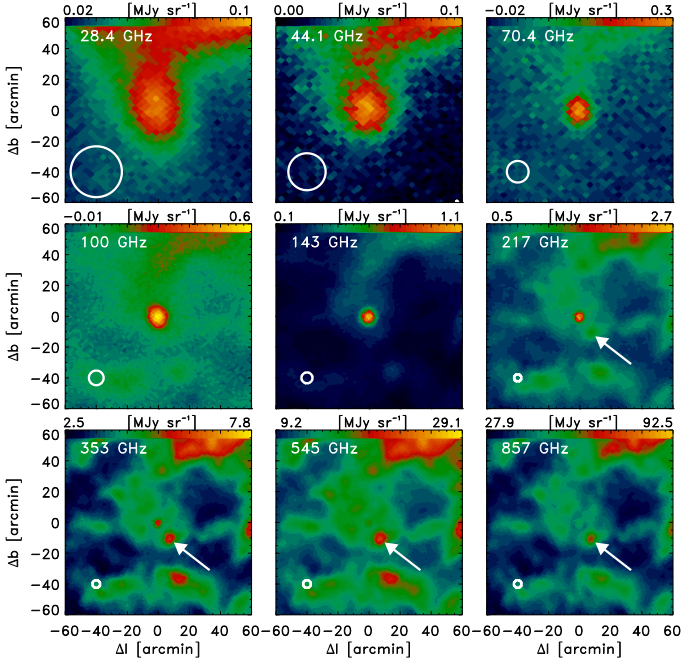}\\
\caption{\Planck\ maps of NGC\,7027. Similar to the case of NGC\,6369, the central component is embedded in an extended dusty halo, visible in most of the HFI channels. There is also a nearby IR source (marked by the arrow), that is not related to the PN. The average FWHM of the effective beam is shown in the bottom-left corner of each map.}
\label{fig:ngc7027}
\end{center}
\end{figure}

\begin{figure}
\centering
\includegraphics[width=88mm]{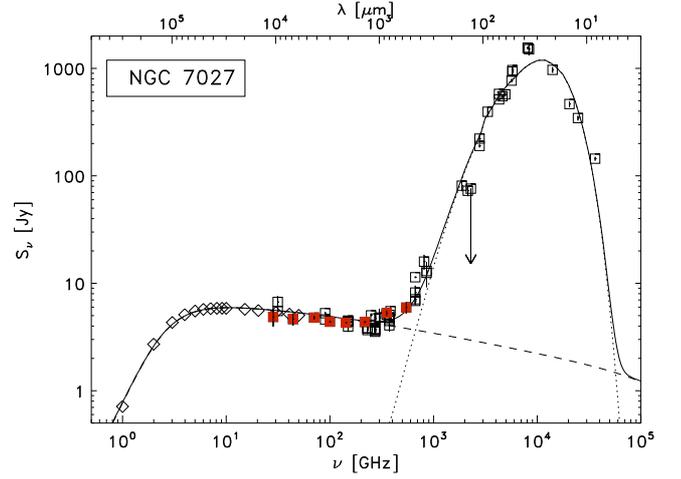}\\
\caption{SED of NGC\,7027. Ancillary measurements, obtained with different instruments, are shown as open squares, while the \Planck\ data are shown as red squares. The arrows indicate upper limits. The diamonds show the radio spectrum of NGC\,7027 given by \citet{zijlstra2008} and evolved to the mean \Planck\ observing time. The continuous line is the model of the SED, obtained by combining both free-free (dashed line) and thermal dust emission (dotted line).}
\label{sp:ngc7027}
\end{figure}

In the case of the Helix, which in the \Planck\ maps appears to be quite extended, we prefer to perform non-blind aperture photometry directly on the \Planck\ maps, assuming an aperture with a radius ($R_\mathrm{a}$) of
\begin{equation}
R_\mathrm{a}= \sqrt{(\mathrm{FWHM})^{2}+(\Omega_\mathrm{})^{2}},
\end{equation}
where $\Omega_\mathrm{}$ is the angular size of the Helix, assumed to be 13\parcm4 \citep{ODell2004}, and the FWHM at each frequency is the average width as given in \citet{planck2011-1.7} and \citet{planck2011-1.6}. The local background and the uncertainty on the aperture flux are estimated in an annulus with a width of $R_\mathrm{ext}=2R_\mathrm{a}$ just outside the aperture. The noise in the map is estimated from the variance in the outer annulus. Uncertainties in the flux estimation are calculated using the sum in quadrature of the rms of the values in the background annulus plus the absolute calibration uncertainties on each map \citep{planck2011-1.1}.

By definition, all the PCCS detections have S/N $\geq4$. However, to be sure of the robustness of our detections, we consider only sources detected in at least three different channels. Because of the specific scientific aim of this paper, namely to build and to model the SED in order to obtain insight on the ionised gas and dust components, we impose the further constraint that sources must be detected by both LFI and HFI. Sources PN\,M1$-$78, PN\,A66\,77, PP\,40 and NGC\,2579, which satisfy the above criteria, were removed from the original sample, since they have been misclassified; they are actually compact \htwo\ regions \citep{Kohoutek2001}. The thermal dust emission of the PN NGC\,6720, the famous Ring Nebula (M\,57), was clearly detected in all the HFI channels, except at 100\GHz. However, due to the low flux level of the free-free emission, it was not detected by LFI. The source, even though it does not meet the above selection criteria, was nevertheless kept in the sample because of the many ancillary radio data points available from the literature that allow the free-free component to be well defined. Our final sample consists of 13 PNe.

\Planck\ flux densities (DETFLUX), as extracted from the PCCS, are reported in Table~\ref{tab:fluxes}, with their associated uncertainties ($1\,\sigma$). For all the detected sources the DETFLUX flux densities are in good agreement with each other and with ancillary observations.

From Table~\ref{tab:fluxes} it is evident that for several sources we do not have detections in all of the \Planck\ channels. This is attributable to a combination of two main effects: PNe are intrinsically weak compared to the sensitivity level of \Planck; and they are located preferentially at low Galactic latitudes, where confusion due to diffuse free-free emission and dust prevents accurate photometry. For example, based on the radio flux densities reported in \citet{Umana2008a} and in \citet{Pazderska2009}, there are five sources, namely NGC\,6302, PN H1$-$12, NGC\,6537, PN M1$-$51, and NGC\,1514, which should have been detected with high S/N, but have only been detected in one or two channels, these possibly being spurious detections. Among them, four are located very close to the Galactic plane, with strong contamination of both diffuse ionised gas and dust. To investigate the missing detection for NGC\,1514, which should have a flux of about 900\,mJy and is located at high Galactic latitude ($b=15\deg$), a direct inspection of the \Planck\ maps was performed. This revealed that the central PN is embedded in a very extended dusty structure preventing a flux measurement (see Fig.\,\ref{fig:ngc1514}). 

In other cases there are hints of extended dusty haloes around the central PN in the higher-frequency \Planck\ maps (see Fig.\,\ref{fig:ngc6369} and Fig.\,\ref{fig:ngc7027}), which, however, are still very well detected in several \Planck\ channels, allowing us to accurately measure fluxes. Usually, the effect of such extended emission is to reduce the S/N. However, in extreme cases, the related flux measurement is disregarded, because from direct inspection of the corresponding \Planck\ map there is strong confusion due to diffuse emission that is not related to the target. As an example, in the case of NGC\,6369 the nearby IR structure (see Fig.\,\ref{fig:ngc6369}) strongly contaminates the source photometry at $\nu \gtrsim 300$\GHz.

To further test the reliability of the \Planck\ data we use the well-known PN NGC\,7027, whose radio flux density evolution has been studied by \citet{zijlstra2008}. These authors determined the rate at which the radio flux density is changing, caused by the expansion of the ionised nebula. The \Planck\ measurements of NGC\,7027,
displayed in Fig.\,\ref{sp:ngc7027} (red squares), are in very good agreement with the radio spectrum provided by \citet{zijlstra2008} when evolved to the mean epoch of the \Planck\ measurements (points pictured as diamond symbols in Fig.\,\ref{sp:ngc7027}), and the sub-millimetre and infrared ancillary data (open squares, \citealp{Telesco1977,Elias1978,Moseley1980,Ulich1981,Gee1984,steppe_etal88,Hoare1992,knapp_etal93,altenhoff_etal94,sandelll94,egan_etal03,murakami_etal07,DiFrancesco2008}).

Before modelling the SEDs, a colour correction has been applied to account for the finite bandpass at each frequency. Since the correction factors depend on the power law of the spectrum inside the band, for each frequency a spectral index has been computed based on the flux density values reported in Table~\ref{tab:fluxes}, relative to the adjacent bands. The colour corrections have then been computed following the procedure described in \citet{planck2013-p02} and \citet{planck2013-p03}. This is typically of the order of a few times 0.1\,\% for LFI channels and up to 7\,\% for HFI channels, and, as such, it is much lower than the uncertainty associated to the flux density. Only one colour correction iteration was applied. The \Planck\ data points of the selected targets are displayed as red squares in Figs.\,\ref{sp:ngc7027}, \ref{fig:SED}, \ref{sp:crl618} and \ref{sp:Helix}.

\section{Physical properties of the source sample}\label{sec:properties}

In the spectral range between IR and radio, the SED of a PN is characterized by two components: the thermal emission from dust grains; and the free-free emission from the ionised part of the CSE. Recently, it has been realised that dust in and around ionised regions in PNe can play an important role in the energetic output of PNe, and up-to-date photo-ionisation codes now include the effect of the dust \citep{VanHoof2000b}.

We have first tested different radiative transfer codes that work in dusty environments to identify the most appropriate for our scientific aim. One such code is the photo-ionisation code {\tt CLOUDY} \citep{Ferland1998}. Although a good match is obtained at near-IR wavelengths, in our trials {\tt CLOUDY} was typically unable to reproduce the long-wavelength spectrum, in particular the optically thick part of the radio continuum, where the spectral slope of the model is typically too steep, implying that radiation transfer is not properly taken into account for the free-free emission. Also, {\tt CLOUDY} aims at reproducing both continuum and line spectra, therefore UV, optical, and IR spectral data are necessary to constrain its output. Since our goal is to fit only the continuum emission, the use of {\tt CLOUDY} would introduce unnecessary degrees of freedom in our modelling. Furthermore, those degrees of freedom would be difficult to constrain with a homogeneous set of spectra for our sources: for example, optical spectra for targeted regions within very extended nebulae would certainly not account for the emission from the whole target, which is the case for the other data from IR surveys and single-dish radio telescopes. We thus prefer to model the thermal dust emission with the publicly-available {\tt DUSTY} code \citep{Ivezic1999} in combination of our own free-free modelling.

We first model the centimetre continuum radiation with our code for free-free emission \citep{Umana2008b} used to fit the SED. This has been improved by implementing a minimization procedure consisting of an iterative process where the free-free emission is calculated within a grid of model parameters, with the ranges and steps gradually restricted until the $\chi ^2$ stabilizes. 
The errors associated to the model parameters have been estimated by varying separately each parameter until the $\chi ^2$ increases of a unit.

Following the wind-shell model, basic geometry usually adopted in the case of the planetary nebulae \citep{taylor_etal87}, each source has been modelled as a central cavity surrounded by a shell of ionised gas. The source geometry is characterized by the linear diameter and the ratio between the internal and external radii ($r_\mathrm{int}$/$r_\mathrm{out}$), where the density distribution of the ionised gas is described by the power law $n_\mathrm{e}(r) \propto r^{-\alpha}$. The radio spectrum has been calculated  with the ratio $r_\mathrm{int}$/$r_\mathrm{out}$, the electron density at $r_\mathrm{int}$ ($n_{\mathrm e}$) and the electron temperature ($T$) as model parameters.

The radius of the free-free source, the distance and the index $\alpha$ describing the spatial distribution of the ionised gas have been set to fixed values to constrain the model. The slope $\alpha$ was set to 2,  consistent with  a spherical wind in steady-state, with the exception of CRL\,618 where it was necessary to vary it to assure a good fit to the observed data (see Table~\ref{tab:ff}). The angular diameters of the radio sources have been mainly derived from the NRAO VLA sky survey \citep[NVSS;][]{Condon1998}. In the case of partially resolved sources, this diameter is related to the actual diameter of the source by a shape factor that accounts for the true structure of the source, and a correction has been applied (following \citealp{VanHoof2000}) to derive the true source size \citep{Umana2008a}.

Distances have been generally adopted from \cite{Pottasch2010}, with exceptions being NGC\,6572 \citep{Phillips2011}, NGC\,6720 \citep{ODell2007}, NGC\,6853 \citep{Stanghellini2008}, NGC\,7009 \citep{fernandez2004}, and NGC\,7027 \citep{zijlstra2008}. 
The best-fit parameter values are summarized in Table~\ref{tab:ff}. The total mass of ionised gas has been calculated numerically by integrating the electron density over the entire structure of the radio source.
 The obtained density and total ionised mass scales with  the adopted distance (D) as $\propto D^{-1/2}$ and as $\propto D^{5/2}$ respectively.

\begin{table*}[tmb]
\begingroup
\newdimen\tblskip \tblskip=5pt
\caption{Source parameters for the free-free emission fit.}
\label{tab:ff}
\nointerlineskip
\vskip -3mm
\footnotesize
\setbox\tablebox=\vbox{
 \newdimen\digitwidth 
 \setbox0=\hbox{\rm 0} 
 \digitwidth=\wd0 
 \catcode`*=\active 
 \def*{\kern\digitwidth}
 \newdimen\signwidth 
 \setbox0=\hbox{+} 
 \signwidth=\wd0 
 \catcode`!=\active 
 \def!{\kern\signwidth}
 \halign{\hbox to 1in{#\leaderfil}\tabskip 0.5em&
 \hfil#\hfil&
 \hfil#\hfil&
 \hfil#\hfil&
 \hfil#\hfil&
 \hfil#\hfil&
 \hfil#\hfil&
 \hfil#\hfil\tabskip 0pt\cr
 \noalign{\doubleline\vskip 2pt}
 \omit&&   &\multispan3 **********Free Parameters********** &&\cr
\cline{4-6}
\omit Source &$r_\mathrm{out}$ & Distance   & $T$**           & {\it r$_\mathrm{int}$/r$_\mathrm{out}$}** & $n_\mathrm{e}^{\dag}$        & $\alpha$\phantom{.0}      & $M_\mathrm{ion}**$ \cr
 \omit        &[arcsec]         & [pc]       & [$10^{4}$\,K]** &                                         & [$10^{4}$\,cm$^{-3}$] &               &[\Msolar]**         \cr
\noalign{\vskip 4pt\hrule\vskip 6pt}
NGC\,6369     & *16.5*          & 1\,200     & 1.0{\s($\pm$0.1)}**           & 0.57{\s($\pm$0.01)}**        & 0.82{\s($\pm$0.02)}                 & 2\phantom{.0} & 0.30{\s($\pm$0.01)}**\cr
NGC\,6572     & **4.8*          & 1\,060     & 0.8{\s($\pm$0.1)}**          & 0.28{\s($\pm$0.01)}**         & 10.3{\s($\pm$0.3)}***               & 2\phantom{.0} & 0.03{\s($\pm$0.01)}**\cr
Helix                & 400.**          & \,*213     & 1.0{\s($\pm$0.2)}**           & 0.59{\s($\pm$0.03)}**        & ****0.0126{\s($\pm$0.0005)}  & 2\phantom{.0} & 0.39{\s($\pm$0.01)}**\cr
NGC\,7009     & *12.**          & \,*860     & 1.0{\s($\pm$0.3)}**           & 0.68{\s($\pm$0.1)}***         & 0.84{\s($\pm$0.04)}                   & 2\phantom{.0} & 0.047{\s($\pm$0.008)}\cr
NGC\,6853     & 170.**          & \,*264     & 0.8{\s($\pm$0.2)}**           & 0.032{\s($\pm$0.001)}       & 2.0{\s($\pm$0.1)}**                  & 2\phantom{.0} & 0.065{\s($\pm$0.006)}\cr
NGC\,6720     & *35.5*          & \,*770     & 0.8{\s($\pm$0.2)}**           & 0.21{\s($\pm$0.01)}**        & 0.47{\s($\pm$0.02)}                   & 2\phantom{.0} & 0.117{\s($\pm$0.007)}\cr
NGC\,6826     & *12.5*          & 1\,200     & 2.2{\s($\pm$0.5)}**           & 0.41{\s($\pm$0.02)**}       & 0.90{\s($\pm$0.04)}                   & 2\phantom{.0} & 0.105{\s($\pm$0.005)}\cr
NGC\,7027     & **4.5*          & \,*980     & 1.7{\s($\pm$0.3)}**           & 0.67{\s($\pm$0.04)}**       & 11.5{\s($\pm$0.3)}***                 & 2\phantom{.0} & 0.053{\s($\pm$0.002)}\cr 
NGC\,6543     & **9.5*          & 1\,000     & 0.9{\s($\pm$0.2)}**           & 0.92{\s($\pm$0.01)}**        & 1.55{\s($\pm$0.06)}                   & 2\phantom{.0} & 0.030{\s($\pm$0.002)}\cr
NGC\,40         & *21.5*          & \,*800     & 1.1{\s($\pm$0.5)}**           & 0.20{\s($\pm$0.01)}**         & 1.27{\s($\pm$0.08)}                   & 2\phantom{.0} & 0.072{\s($\pm$0.006)}\cr
CRL\,618       & **0.45          & \,*900     & 0.36{\s($\pm$0.05)}       & 0.021{\s($\pm$0.001)}          & 9\,800{\s($\pm$500)}******       & 1.3                     & ***0.00082{\s($\pm$0.00004)}\cr
IC\,418            & **6.0*          & 1\,000     & 1.3{\s($\pm$0.3)}**           & 0.44{\s($\pm$0.02)}**        & 5.2{\s($\pm$0.3)}**                     & 2\phantom{.0} & 0.045{\s($\pm$0.003)}\cr
NGC\,3242     & *15.5*          & \,*550     & 0.9{\s($\pm$0.2)}**           & 0.7{\s($\pm$0.1)}****         & 0.77{\s($\pm$0.05)}                    & 2\phantom{.0} & 0.024{\s($\pm$0.002)}\cr
\noalign{\vskip 3pt\hrule\vskip 4pt}
}}
\endPlancktablewide
\vspace{-2mm}
\begin{list}{}{}
\item[] {{$^{\dag}$ Electron density at the inner radius.}}
\end{list}

\endgroup

\end{table*}

\begin{table*}[tmb]
\begingroup
\newdimen\tblskip \tblskip=5pt
\caption{Source parameters for the {\tt DUSTY} fit. am-C indicates amorphous carbon \citep{hanner1988}, Sil-Ow indicate O-deficient silicates \citep{Ossenkopf1992}, and Sil-DL are astronomical silicates \citep{draine1984}.}
\label{tab:dust}
\nointerlineskip
\vskip -3mm
\footnotesize
\setbox\tablebox=\vbox{
 \newdimen\digitwidth 
 \setbox0=\hbox{\rm 0} 
 \digitwidth=\wd0 
 \catcode`*=\active 
 \def*{\kern\digitwidth}
 \newdimen\signwidth 
 \setbox0=\hbox{+} 
 \signwidth=\wd0 
 \catcode`!=\active 
 \def!{\kern\signwidth}
 \halign{\hbox to 1in{#\leaderfil}\tabskip 0.5em&
 \hfil#\hfil&
 \hfil#\hfil&
 \hfil#\hfil&
 \hfil#\hfil&
 \hfil#\hfil&
 \hfil#\hfil&
 \hfil#\hfil&
 \hfil#\hfil&
 \hfil#\hfil\tabskip 0pt\cr
 \noalign{\doubleline\vskip 2pt}
\omit&&&&   &\multispan3 Free Parameters &&\cr
\cline{6-8}
\omit Source& $R_\mathrm{int}$& $R_\mathrm{int}$/$R_\mathrm{c}$& Luminosity& $T_\mathrm{star}$& $T_\mathrm{dust}$& $R_\mathrm{out}$/$R_\mathrm{int} ^{\dag}$& $\tau_{V}$&Grains& $M_\mathrm{dust}$\cr
\omit& [arcsec]& [$10^{4}$]& [$10^3\,\Lsolar$]& [K]& [K]&&& composition& [$10^{-4}$\,\Msolar]\cr
\noalign{\vskip 4pt\hrule\vskip 6pt}
%
NGC\,6369 &   2.8{\s($\pm$0.2)}  &   \,200{\s($\pm$20)}   &  3**   &  70\,000 & 140{\s($\pm$4)} &   **100   &  0.030{\s($\pm$0.001)}  & am-C   & 4{\s($\pm$3)} \cr
NGC\,6572 &   3.7{\s($\pm$0.3)}  &   \,180{\s($\pm$20)}   &  8.3  &  80\,000 & 160{\s($\pm$5)} &    ***30   &  0.013{\s($\pm$0.001)}  & am-C   &   **0.7{\s($\pm$0.3)}\cr
Helix     &   140{\s($\pm$20)}****   & 22\,000{\s($\pm$4000)}  & *0.12 & 110\,000* &  *34{\s($\pm$2)} &  ***20   &  **0.0085{\s($\pm$0.0001)} & am-C   & 18{\s($\pm$7)}* \cr
NGC\,7009 &   6.0{\s($\pm$0.4)}  &   \,300{\s($\pm$20)}    &  5**    &  80\,000 & 115{\s($\pm$3)} &   **100   &  **0.0029{\s($\pm$0.0001)} & Sil-DL &  2{\s($\pm$1)}\cr
NGC\,6720 &   4.8{\s($\pm$0.4)}  &  2\,400{\s($\pm$200)}   &  0.2  & 120\,000* &  *80{\s($\pm$3)} &   **200   &  0.035{\s($\pm$0.002)}  & am-C   & 11{\s($\pm$5)}*\cr
NGC\,6826 &   2.3{\s($\pm$0.2)}  &  \,100{\s($\pm$10)}    &  *1.64 &  47\,500 & 120{\s($\pm$5)} &    ***30   &  0.030{\s($\pm$0.002)}  & Sil-DL & ***1.5{\s($\pm$0.6)}\cr
NGC\,7027 &   2.6{\s($\pm$0.1)}  &   \,580{\s($\pm$20)}    & 10***    & 190\,000* & 190{\s($\pm$2)} &   **200   &  0.025{\s($\pm$0.004)}  & am-C   & 4{\s($\pm$2)}\cr
NGC\,6543 &   4.8{\s($\pm$0.5)}  &   \,130{\s($\pm$10)}    &  5.2  &  56\,000 & 120{\s($\pm$5)} &   **100   &  0.012{\s($\pm$0.001)}  & Sil-DL & 6{\s($\pm$3)}\cr
NGC\,40   &   2.3{\s($\pm$0.2)}  &       \,74{\s($\pm$6)}     &  1.7  &  50\,000 & 160{\s($\pm$5)} &   **100   &  0.040{\s($\pm$0.002)}  & am-C   & 2{\s($\pm$1)}\cr
CRL\,618  &  **0.29{\s($\pm$0.03)} &  ****2.0{\s($\pm$0.2)}   &  6**    &  30\,000 & 450{\s($\pm$4)} & 10\,000 & 50{\s($\pm$4)}*******  & am-C   & 4\,000{\s($\pm$2\,000)}*\cr
IC\,418   &   2.5{\s($\pm$0.3)}   &  \,20{\s($\pm$2)}     & 10***    &  35\,000 & 200{\s($\pm$8)} &   **200   &  0.040{\s($\pm$0.001)} & am-C   & 6{\s($\pm$3)}\cr
NGC\,3242 &   0.8{\s($\pm$0.2)}   &   *80{\s($\pm$20)}    &  0.6  &  80\,000 & 200{\s($\pm$20)}&  **100   &  0.007{\s($\pm$0.001)} & Sil-Ow & ***0.3{\s($\pm$0.2)}\cr
\noalign{\vskip 3pt\hrule\vskip 4pt}
}}
\endPlancktablewide
\vspace{-2mm}
\begin{list}{}{}
\item[] {{$^{\dag}$ To estimate the errors on the input and output parameters the
{\tt DUSTY} code has been used by varying arbitrarily this parameter by a factor of two.}\\
{$^{\dag \dag}$ $R_c$ indicates the radius of the central star.}
}
\end{list}

\endgroup
\end{table*}

{\tt DUSTY} modelling was performed after the best-fit value for the radio emission was obtained. General inputs for {\tt DUSTY} are: the central object radiation field (assumed to be a Planck function with an appropriate temperature and luminosity); composition and size distribution of the dust grains; and a model for the dusty shell (density distribution, optical thickness $\tau_V$ in the visual, geometrical thickness, and temperature of the grains at the inner radius of the shell, \textbf{$T_{dust}$}). {\tt DUSTY} assumes spherical symmetry for the dusty envelope. Although such symmetry is not always observed in our targets, {\tt DUSTY} can still be used to obtain first-order approximations of nebular parameters.

{\tt DUSTY} allows us to use six  different grain types: warm O-deficient silicates (Sil-Ow); cold O-rich silicates (Sil-Oc); astronomical silicates (Sil-DL); graphite (grf-DL); amorphous carbon (am-C); and silicon carbide (SiC-Pg). To assess the dominant mineralogy of the dust in the circumstellar envelope of our targets (i.e., C-rich versus O-rich), we use the direct information on dust features provided by \ISO\ spectra, namely: \citet{Cernicharo2001} for CRL\,618; \citet{Bernard-Salas2001} for NGC\,7027; \citet{Bernard-Salas2005} for NGC\,6543; \citet{Volk2003} for IC\,418; \citet{Szczerba2001} for NGC\,6369 and NGC\,40; \citet{Surendiranath2008} for NGC\,6826; and \citet{Phillips2010} for NGC\,7009. In the remaining cases, we use the C/O gas-phase abundance \citep{liu2001}, as this is a good indicator of the presence or absence of Carbon-based dust grains (i.e., PAH) in several PNe \citep{Cohen2005}. Furthermore, the choice among different kinds of silicates or Carbon-based grains was based on what provides the best fit to the SED. The standard MRN \citep*{Mathis1977} grain size distribution (grain radius $a = 0.005$--0.25\um, with a distribution following $a^{-3.5}$) was applied to all sources. 

The density distribution was set to depend on $r^{-2}$ for all sources. Such a distribution follows naturally from the assumption that the circumstellar material was ejected during the AGB phase at a constant mass-loss rate. The only exception is NGC\,3242, where a density distribution of $r^{-1.1}$ was necessary to fit the dust contribution. NGC\,3242 is not particularly different from the other targets with respect to the optical morphology, evolution, and position, and therefore it is very difficult to justify a different dust distribution in the model. However in this context, it may be worth pointing out that the presence of a system of concentric rings, whose brightness is higher in the mid-IR than in the optical, located in an extended halo around the source, has been recently reported by \citet{Phillips2009}. 
Furthermore,  from the analysis of its HST images, \citet{ruiz11}
modelled its H$\alpha$ brightness spatial distribution
as a thin shell with a constant density 
embedded within an outer shell, characterized by a radial density profile much shallower than the classical profile for a free-expanding wind at a constant mass-loss rate ($r^{-2}$).
This is in agreement with our dust modelling for NGC\,3242, whose density profile is less steep than that of the other PNe.
Interestingly, the radius of the dusty CSE as derived from our {\tt DUSTY} modelling (about 80\arcs) matches quite well with the reported size of the extended halo (about 75\arcs) where the concentric dusty rings are located.

We have observed in NGC\,7009 and NGC\,6826 that the fit underestimates the near- and mid-IR emission, despite taking into account the free-free emission from our radio model. This may point to the presence of a separate population of hot grains ($T_\mathrm{dust} \ge 500$ K). Very little is known about the spatial distribution of dust in PNe. However, given the variety of shapes of PNe observed in the optical and near-IR, the presence of a hotter population of grains may be linked to the specific distribution of dust, with some warmer regions closer to the central star. It is difficult to estimate the amount of such an extra dust component without adding further free parameters to the {\tt DUSTY} fit. On the other hand, even if such a warm dust component exists, our simulations indicate that it does not have any effects on the temperature of the cooler dust component traced by the millimetre and far-IR emission. The slope of the SED at long wavelengths is determined by the opacity and the size of the dust shell (i.e., the contribution from colder dust). We produced our models taking into account the necessity to properly fit the \Planck\ data and imposing that the outer temperature of the dust shell must match a mean ISM temperature of 10--20\,K. 

\begin{figure*}
\centering
\includegraphics[width=180mm]{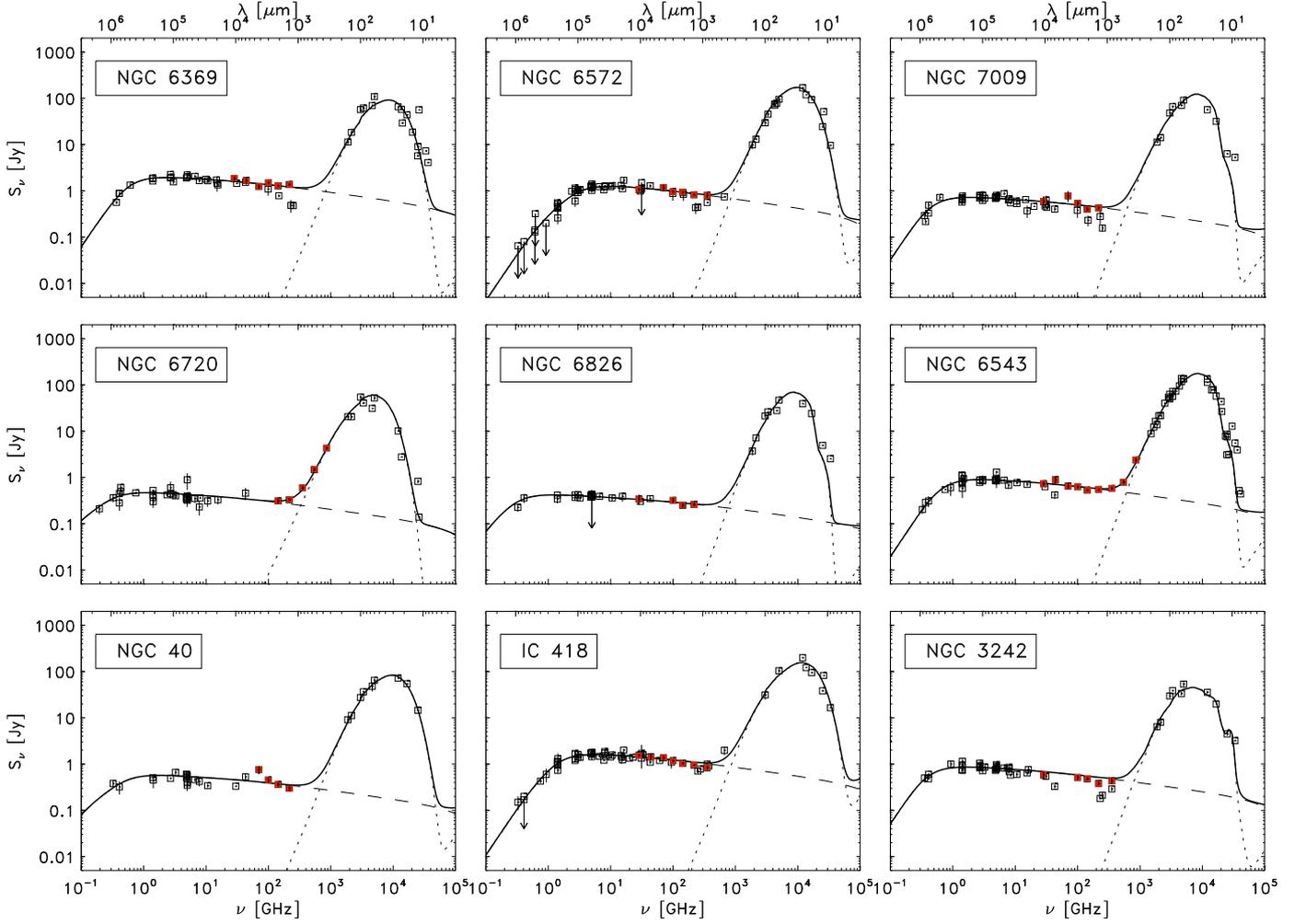}\\
\caption{Models of the SEDs for our sample of Galactic PNe. 
The continuous line is the combination of both free-free (dashed line) and thermal dust emission (dotted line).
Measurements obtained with different instruments and collated from the literature are shown as open squares. \Planck\ measurements are indicated by red squares. The arrows indicate upper limits.}
\label{fig:SED}
\end{figure*}

{\tt DUSTY} treats the problem of radiation transfer independently from the distance to the star by taking advantage of the scaling properties of the radiative transfer equation \citep{Ivezic1997}. This implies that its output must be scaled to the actual data. We calculated the scaling factor taking into account the estimates of distance and luminosity available in the literature \citep{goodrich1991,sabbadin2004,prinja2007,gruenwald2007,Surendiranath2008,zijlstra2008,pottasch2008,morisset2009,Pottasch2010,vanhoof2010,monteiro2011}. Adjustments of both parameters (distance and luminosity) were necessary to match the observational points. From the derived parameters, it is possible to obtain an estimate of the total mass of dust in the shell from the equations derived by \citet{Sarkar2006}:
\begin{equation} 
M_\mathrm{dust}=4\, \pi \,R_\mathrm{int}^{2} \,Y\, \left( \frac{\tau_{500}}{\kappa_{500}} \right),
\end{equation}
which is valid in the case of an $r^{-2}$ density distribution and 
\begin{equation}
M_\mathrm{dust} = \frac{4 \pi \tau_{500}}{19 \kappa^{}_{500}}R^2_\mathrm{int} \frac{Y^{1.9}-1}{1-Y^{-0.1}},
\end{equation} 
which is valid for NGC\,3242, where the relation $\rho \propto r^{-1.1}$ is required to model the observational data. $Y=R_\mathrm{out}/R_\mathrm{int}$ is the relative shell thickness, $\tau_{500}$ is the shell optical depth at 500\um\ (which is an output parameter from {\tt DUSTY}) and $\kappa_{500}$ is the dust mass absorption coefficient at 500\um. $R_{int}$ and $R_{out}$ are here the inner and outer radii of the dusty region, respectively, which do not necessarily match with the inner and outer radii of the ionised shell.

The optical properties of a material can be described with the use of  a complex refractive index $m_\lambda=\tilde{n}_\lambda + \imath \tilde{k}_\lambda$, where $\tilde{n}_\lambda=\frac{ck}{\omega}$ and $\tilde{k}_\lambda = \frac{c\alpha_{abs}}{2\omega}$, with $k$ the wave number, $c$ speed of light, $\alpha_{abs}$ the absorption coefficient, and $\omega$ the frequency times $2\pi$. 
From  the chosen set of optical constants (\citealp{Ossenkopf1992}, \citealp{draine1984} or \citealp{hanner1988}),
we can then calculate the dust absorption coefficient. The dust extinction cross section is defined as $\sigma_\mathrm{ext}=\pi a^2 Q_\mathrm{ext}$, where $a$ is the grain radius and $Q_\mathrm{ext}$ the extinction efficiency. At sub-mm wavelengths, scattering can be neglected and the extinction is due to absorption only:
\begin{equation}
\sigma_\mathrm{abs} = \pi a^3 \frac{Q_\mathrm{abs}}{a}.
\end{equation}
The absorption coefficient per unit volume of dust is defined as 
\begin{equation}
\kappa^{}_V = \frac{\sigma_\mathrm{abs}}{V},
\end{equation}
where $V$ is the average volume of a dust grain. Hence
\begin{equation}
\kappa^{}_V=\frac{\sigma_\mathrm{abs}}{\frac{4}{3}\pi a^3}=\frac{3}{4}\frac{Q_\mathrm{abs}}{a}.
\end{equation}

An approximation of $Q_\mathrm{abs}/a$ as a function of only $\lambda$ can then be found, such that \citep{Andersen1999}
\begin{equation}
\kappa^{}_V=\frac{6\pi}{\lambda}\,\mathrm{Im}\left[\frac{m_\lambda^2-1}{m_\lambda^2+2}\right].
\end{equation}
Once $\kappa^{}_V$ is known, we assume typical values of dust grain density $\rho_\mathrm{g}$ of about 3\,g\,cm$^{-3}$ for silicate grains and 2\,g\,cm$^{-3}$ for carbon grains and then calculate the absorption coefficient per unit mass $\kappa^{}_M={\kappa_V} / {\rho_\mathrm{g}}$, which for our sets of optical constants gives us values of $\kappa_{500}$ of 0.98, 1.34, and 2.07\,cm$^{2}$\,g$^{-1}$, as appropriate for the kinds of silicates and carbon dust adopted in our fits, namely Sil-Ow \citep{Ossenkopf1992}, Sil-DL  \citep{draine1984}, and am-C \citep{hanner1988}, respectively.

The relevant parameters of the modelling are summarized in Table~\ref{tab:dust}, where we report the inner radius, the ratio between the inner radius of the envelope to the radius of the central object ($R_\mathrm{c}$), the central star luminosity, the temperature of the dust at the inner radius, the relative thickness of the envelope, ($R_\mathrm{out}$/$R_\mathrm{int}$), the optical depth of the envelope at 0.5\um\ ($\tau_{V}$), the dust composition, and the derived dust mass. 
The free parameters, given as input of {\tt DUSTY}, are: $T_\mathrm{dust}$; $R_\mathrm{out}$/$R_\mathrm{int}$; 
and $\tau_{V}$, the other parameters reported in Table~\ref{tab:dust} are model outputs or fixed parameters. We should stress here that {\tt DUSTY} does not have a minimization procedure, therefore the best combination of parameters is obtained through an iterative process consisting of different trials until a good match with the measurements is reached. For NGC\,6853 we only report values from the free-free modelling, since no reliable IR ancillary data---necessary to constrain the modelling of the dust contribution---were available. 

One major caveat of using {\tt DUSTY} is that the model considers only a central heating source, neglecting a possible contribution due to the external heating by the interstellar radiation field (ISRF). However, {\tt DUSTY} allows the modelling of the UV field due to the central star at each point of the circumstellar shell. This field, diluted at the outermost part of the shell, is 4--5 orders of magnitude higher than the value of the ISRF, which is assumed to be $1.6 \times 10^{-3}$ erg cm$^{-2}$ s$^{-1}$. We thus conclude that the effect of the ISRF is negligible for most of the PNe in our sample. One exception is CRL\,618, whose dust mass could be overestimated if the ISRF is excluded, because the optical depth of its shell is very high (see Table~\ref{tab:dust}).

As previously mentioned, the {\tt DUSTY} code assumes spherical symmetry for the dusty envelope.
To test the reliability of the physical parameters estimated 
under this assumption, we look for new sets of parameters that reproduce the observed SEDs, after setting the ratio of outer to inner radius to the double and to half of our best value for it. This would then tell us how the physical parameters would be affected by assuming that the size of the modelled sphere were twice our best value or half of it. 
We thus test how the geometry of the source influences the fit and derive ranges of values
for the other input ($T_\mathrm{dust}$ and $\tau_{V}$) and output 
(respectively: the inner radius-$R_\mathrm{int}$, the ratio of the inner to the stellar radius-$R_\mathrm{int}$/$R_\mathrm{c}$,
and the mass of dust-$M_\mathrm{dust}$) parameters.
%
%
As expected, for the output parameter $M_\mathrm{dust}$ we have found a variation of about 50\%,
in agreement with the adopted range of shell thickness. 
The ranges of  the other model parameters thus estimated are instead not very large: $R_\mathrm{int}$, $R_\mathrm{int}$/$R_\mathrm{c}$ and $\tau_{V}$ vary within about 10\%, and $T_\mathrm{dust}$ less than 5\%, on average. This confirms the goodness of our nebular parameters as a first-order estimation. 
These ranges are reported in Table~\ref{tab:dust}
as errors associated to the {\tt DUSTY} parameters.

Results for the SEDs, considering both thermal dust and ionised gas contributions, are shown in Fig.\,\ref{sp:ngc7027} and Fig.\,\ref{fig:SED}. 
 
\Planck\ measurements have provided good spectral coverage of the PN SEDs, in particular at those frequencies where free-free and thermal dust emission contributions overlap. Some authors \citep{Hoare1992} have investigated the possible presence of cool ($\sim$20\,K) dust in the CSEs of PNe. The presence of such a component can be excluded for NGC\,6720 and NGC\,6543 directly from \Planck\ data, which nicely follow the combination between free-free and thermal dust emission. However, this result can be extended to the whole sample, since in all cases the SEDs can be explained in terms of free-free and warm thermal dust emission only.

For three sources, namely NGC\,6369, NGC\,7009, and NGC\,3242, the high frequency \Planck\ measurements appear to be systematically higher than the ancillary data. However, all the millimetre ancillary data for those sources are from the same study \citep{casassus_etal07}, where 31\GHz\ and 230 plus 250\GHz\ fluxes are provided, which may suffer from some effects due to the fact that the dusty envelope is partially resolved. Supporting this conclusion, recent LABOCA/APEX measurements (Cerrigone et al. in prep.) of NGC\,7009 and IC\,418 are in good agreement with those obtained from \Planck.

Finally, one of our sources, NGC\,40, appears to have a higher emission with respect to the model at around 70\GHz. With the current data it is difficult to estimate whether this is related to background contamination and/or to an extra contribution (i.e. anomalous emission). It is worthwhile noting that the presence of strong PAH emission has been reported in NGC\,40 \citep{Ramos2011}, which might be linked to radiation of very small spinning dust grains.

Typical densities and temperatures in the ionised regions are in agreement with those that are generally observed \citep{Buckley1995}. The derived values of total ionised masses are also in agreement with values reported in the literature, which are of the order of a few times 0.01\,\Msolar. Besides the extreme cases of CRL\,618 and the Helix, which will be analysed separately in the following sections, the dust temperatures obtained are in the range 80--200\,K (Table~\ref{tab:dust}), in agreement with previous works. In particular, very close values have been obtained for NGC\,40 \citep{Ramos2011} and for NGC\,7009 \citep{Phillips2010} from the analysis of \Spitzer\ data. We also confirm the general evolutionary trend in grain temperature observed by other authors on the basis of IR data \citep{Phillips2011}, where more evolved PNe (i.e., NGC\,6543 and NGC\,6720) display lower values. In general, the inner radius of the dusty envelope appears to be coincident with or located inside the ionised region. This leads to the conclusion that ionised gas and dust spatially coexist in these sources, despite the different physical conditions in which these components are presumed to survive. This points to the presence of a possible shielding effect, allowing the dust to survive in the harsh environment of the UV radiation field of the central star.

The total dust mass, as derived from our modelling, is in the range of values reported by \citet{Gurzadyan1997}, and, in general, two orders of magnitude smaller than that of the ionised gas mass, but it is distributed in a much more extended spatial region. However, such values of $M_\mathrm{dust}$ have to be considered as lower limits. The data collected in this paper mainly trace dust at around 100 K while our modelling pointed out in several cases the possible presence of an extra component of warmer dust, whose contribution cannot be estimated from this data set.

\section{The proto planetary nebula (PPN) CRL\,618} \label{sec:crl618}

One of our targets is CRL\,618, a PPN that entered the post-AGB stage about 200 years ago and is rapidly evolving towards a PN. It provides a unique opportunity to study the physical processes taking place immediately before the birth of a PN. Multi-wavelength, multi-epoch observations (IR, radio, millimetre, and optical) have provided us with a complex picture of the source  \citep{Sanchez2004}, consisting of:
\begin{enumerate}[1)]
\item a torus-like dusty structure located in the equatorial plane and obscuring the central object;
\item two shock-excited optical lobes, extending in the polar axis of the star;
\item a molecular envelope, composed of the AGB remnant CSE plus a fast bipolar outflow;
\item a compact \htwo\ region close to the central object, indicating the onset of ionisation of the envelope.
\end{enumerate}

Although numerous observational and theoretical efforts have been made to understand CRL\,618, several interesting puzzles remain. Among them is the free-free emission from the central \htwo\ region, whose variability has been reported by several authors \citep{Kwok1981, Sanchez2004b, Wyrowski2003}. This somewhat controversial behaviour, implyies that CRL\,618 exhibits the activity of a post-AGB wind, which is poorly understood \citep{Garcia2005}.

 
In order to study the evolution of the radio emission, we have analysed multi-frequency radio data obtained in different epochs and we have concluded that the increase of the radio emission, the derived ionised mass, and the size of the radio nebula are consistent with the hypothesis \citep{Kwok1984} of the expansion of an ionisation front caused by an increase in central star temperature and a larger output of UV photons. Similar conclusions were independently achieved by \citet{tafoya}. 
 
In this framework, it is evident how important it is to build the observed SED from almost coeval measurements for this strongly variable source, especially in the millimetre range, where the strongest variability has been reported \citep{Sanchez2004a}. \Planck\ provides for the first time a coverage of the source spectrum in a very poorly explored spectral region, allowing us to derive information on the physical properties of the associated circumstellar envelope without suffering any variability effects.

\begin{figure}
\centering
\includegraphics[width=88mm]{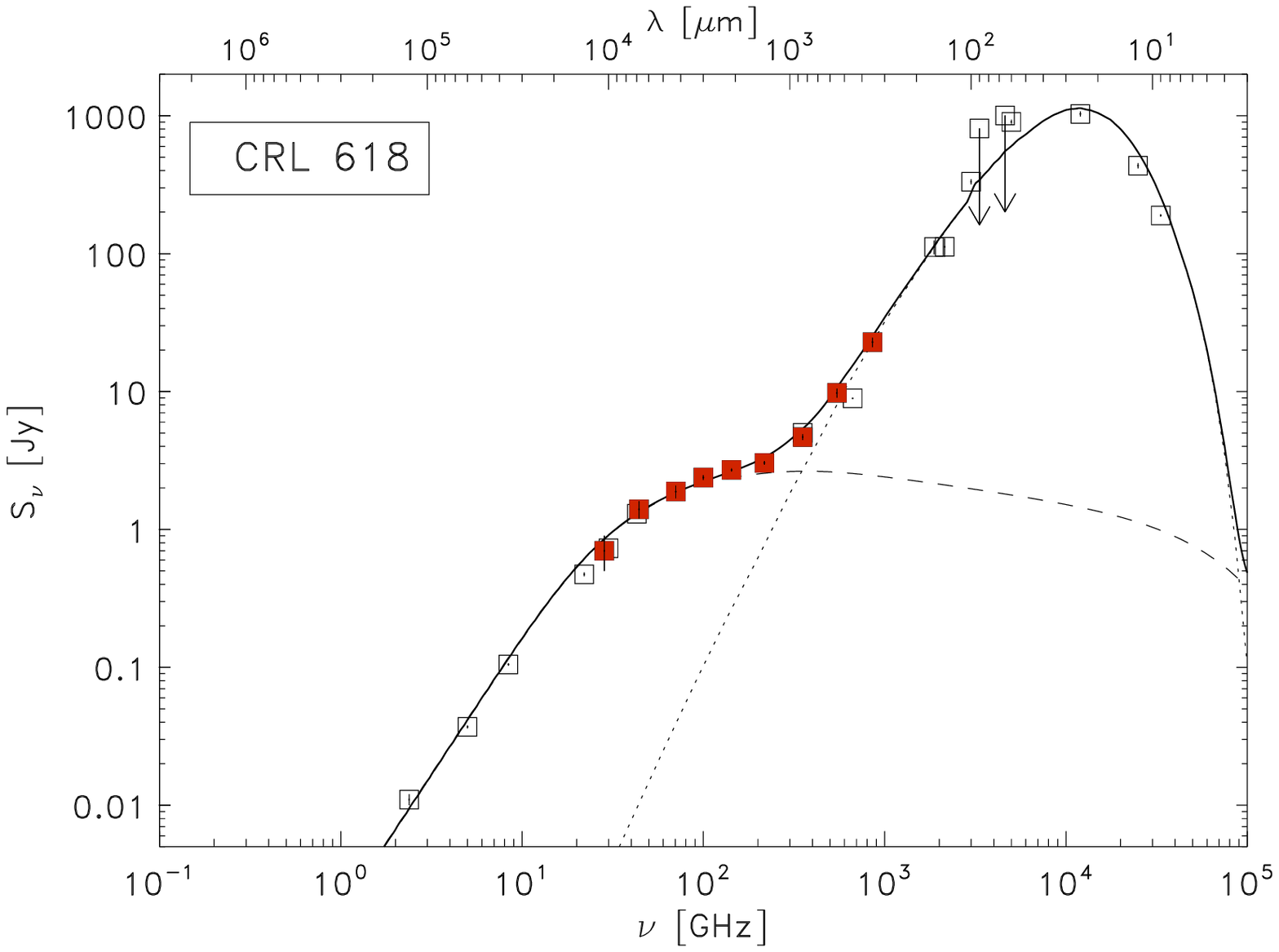}\\
\caption{Model of the SED of CRL 618 (continuous line) considering both free-free (dashed line) and thermal dust emission (dotted line).
Ancillary measurements, obtained with different instruments
(\IRAS, \citealp{helou_walker88}; \AKARI, \citealp{murakami_etal07}),
are shown as open squares, while \Planck\ data are shown as red squares. 
The arrows indicate upper limits.
For the ancillary radio measurements ($\nu \leq$ 40\GHz) only VLA data coeval to \Planck\ observations have been used (see text).}
\label{sp:crl618}
\end{figure}

\begin{table}[tmb]
\begingroup
\newdimen\tblskip \tblskip=5pt
\caption{Continuum flux density for CRL\,618 from the CALSUR program (2009-VLA D configuration).}
\label{tab:radio}
\nointerlineskip
\vskip -3mm
\footnotesize
\setbox\tablebox=\vbox{
 \newdimen\digitwidth 
 \setbox0=\hbox{\rm 0} 
 \digitwidth=\wd0 
 \catcode`*=\active 
 \def*{\kern\digitwidth}
 \newdimen\signwidth 
 \setbox0=\hbox{+} 
 \signwidth=\wd0 
 \catcode`!=\active 
 \def!{\kern\signwidth}
 \halign{\hbox to 1in{#\leaderfil}\tabskip 0.5em&
 \hfil#\hfil&
 \hfil#\hfil&
 \hfil#\hfil&
 \hfil#\hfil\tabskip 0pt\cr
 \noalign{\doubleline\vskip 2pt}
Frequency& 5\,GHz& 8.3\,GHz& 22\,GHz& 43\,GHz\cr
&(C-band)& (X-band)& (K-band)& (Q-band)\cr
\noalign{\vskip 4pt\hrule\vskip 6pt}
Flux density [mJy]& 37$\pm$0.9& 105$\pm$2.0& 473$\pm$15& 1300$\pm$40\cr
\noalign{\vskip 3pt\hrule\vskip 4pt}
}}
\endPlancktable
\endgroup
\end{table}

Assuming that the variability is mainly related to the evolution of the central \htwo\ region, and therefore that it will mostly influence the free-free contribution, we have retrieved from the Karl G. Jansky Very Large Array (VLA) archive the data sets observed as close in time as possible to the nominal \Planck\ mission, defined as 12 August 2009 through 27 November 2010. We found multi-frequency, D-configuration data (Program: CALSUR), obtained in November 2009. Data reduction and imaging were performed by using the software package {\tt CASA} 3.0.0, following standard procedures. The derived flux density and related rms noise are summarized for each frequency in Table~\ref{tab:radio}. \citet{Sanchez2004a} reports strong variability in the millimetre range. Although continuous monitoring of the millimetric flux density would be necessary for a full characterization of its behaviour, the authors concluded that the millimetric light curve is more complicated than a steady monotonic function and consists of a combination of epochs with increasing and decreasing flux. Such variations have a characteristic time scale of the order of 2--3 years \citep{Sanchez2004a}. On the other hand, from \citet{Umana2013}, a steady increase of the flux density is observed between about 1 and 20\GHz\, over a large temporal baseline spanning more than 20 years. At 15\GHz\ an increase of around 30\,\% is observed in 10 years. Therefore, even if the VLA and \Planck\ measurements are not exactly simultaneous, we do not expect any significant variation over the one year time between them. This conclusion is also supported by the agreement between the 43\GHz\ VLA measurement (Table~\ref{tab:radio}) and the 44\GHz\ \Planck\ measurement (Table~\ref{tab:fluxes}).

To model the SED, we followed the same procedure as described in the previous section. The size of the radio source was set by following the results from \citet{Umana2013}, who measured a total extension of $0\parcs9$. The results of our modelling of both the free-free and thermal dust emission are shown in Fig.\,\ref{sp:crl618}, where a clear thermal dust contribution is superimposed on a typical free-free spectrum with a turnover frequency of around 40\GHz. The assumed gradient in the density distribution, necessary to fit the observed data, is consistent with the stellar wind model of \citet{Martin1988}, even though the index of the power law describing the density distribution ($\alpha=1.3$)
indicates we are in the presence of something different from an isotropic canonical stellar-wind \citep{taylor_etal87}. In the optically thin part of the spectrum, an increasing contribution from thermal dust is evident. This is consistent with results reported by \citet{Nakashima2007}, whose SMA observations pointed out the presence at 690\GHz\ of an extended component partially resolved by the $1\parcs0 \times 0\parcs8$ synthetic beam of the interferometer and  therefore morphologically different from the compact free-free component. All the relevant parameters of the free-free and {\tt DUSTY} fits are summarized in Table~\ref{tab:ff} and in Table~\ref{tab:dust}.

The physical characteristics of the envelope surrounding CRL\,618 stand out with respect to those of the other targets. This is consistent with CRL\,618 being a very compact (high $n_\mathrm{e}$) young PN (or a PPN), where the ionisation has just started (a very small value for $M_\mathrm{ion}$). The compact PPN is still embedded in an extended dusty environment---a remnant of the previous evolutionary stage---containing at least 0.4 \Msolar\ of dust. This value of the dust mass in the CSE of CRL\,618 is comparable to that derived for the mass of the molecular gas \citep{Sanchez2004a}, estimated as 0.7 \Msolar.

\section{The Helix} \label{sec:helix}

\subsection{Overall picture}

NGC\,7293, the Helix, is one of the closest (213\,pc; \citealp{harris1997}) bright and evolved PNe. Its central star is well resolved from the surrounding nebula and is already on the white dwarf (WD) cooling track, with a temperature of about  $1.1 \times 10^{5}$\,K. Thanks to its proximity, several authors have studied the Helix in different spectral ranges and at different spatial scales \citep{Speck2002,ODell2004,Hora2006, meaburn_etal13}. All together, these observations have provided us with a detailed picture of this evolved nebula, revealing both its highly inhomogeneous nature and its overall large-scale structure. The main ring consists of at least two different emission components with a highly structured texture at a separation of about $4\parcm2$---the so-called cometary knots. An outer ring extends up to 25\arcm\ from the central star. The outer ring/disk is flattened on the side colliding with the ambient interstellar medium. For a complete census of all the structures identified in the Helix see \citet{ODell2004}. Evidence for extended emission was also reported by \citet{Speck2002} in both $H_{\alpha}$ and FIR, indicating the presence of a large halo consisting of ionised gas and dust.
Such large halo is clearly visible also in the ultraviolet image of the Helix obtained with the \GALEX\ satellite
\citep{Zhang2012,meaburn_etal13}. Furthermore, the Helix image performed at  ultraviolet wavelengths reveals the existence of a very outer and faint extended halo (diameter up to $40\arcm$), bow shaped on the east side and jet-like shaped on the west side. These features are oriented along the proper motion direction and were firstly identified in Helix images obtained in the $H_\alpha$ and [\ion{N}{ii}] 6584 \AA\ emission lines \citep{meaburn_etal05}.

\begin{figure}
\centering
\includegraphics[width=88mm]{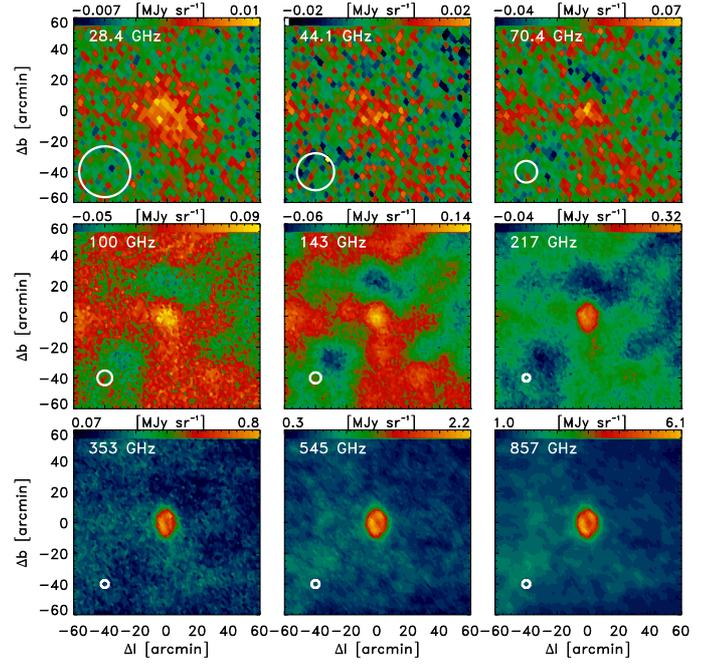}
\caption{\Planck\ maps of the Helix (in Galactic coordinates). At each frequency, the corresponding beam (FWHM) is indicated in the bottom-left corner. In addition to the well known structures, associated with the evolved PN, diffuse emission is also present up to one degree from the central object. }
\label{HELIX}
\end{figure}

\begin{figure}
\centering
\includegraphics[width=88mm]{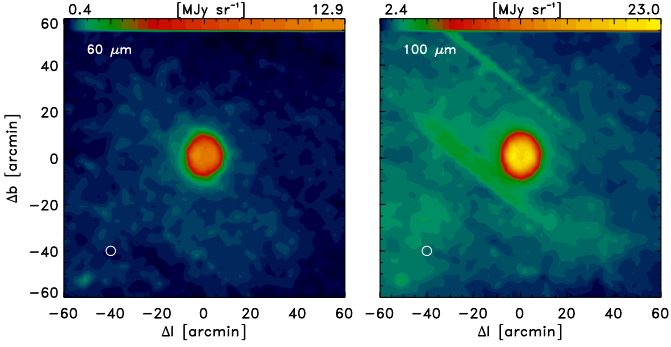}
\caption{IRIS 60 and 100\um\ maps of the Helix (in Galactic coordinates). At each frequency, the corresponding beam (FWHM) is indicated in the bottom-left corner. }
\label{HELIX_IRIS}
\end{figure}

\begin{figure}
\centering
\includegraphics[width=88mm]{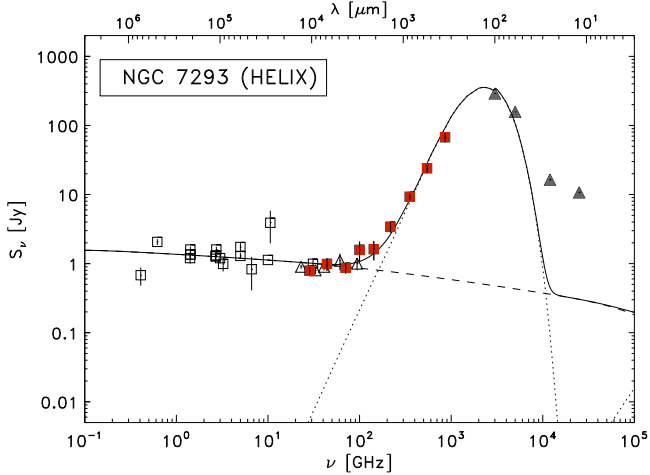}
\caption{Model of the SED of NGC 7293 (continuous line) considering both free-free (dashed line) and thermal dust emission (dotted line).
Ancillary radio measurements, obtained with different instruments, are shown as open squares
(reference numbers given in Table~\ref{tab:fluxes}: 2, 3, 6, 7, 16, 17, 18, 19, 23, 29 and 52), \WMAP\ 7-year data as open triangles, aperture photometry from IRIS as filled triangles, and \Planck\ data are shown as red squares. IRIS data at 12\um\ and 25\um, which are contaminated by emission lines, are also shown in the plot for completeness, but were not considered in the fitting procedure (see text).}
\label{sp:Helix}
\end{figure}

The Helix was detected in all \Planck\ channels. The LFI channels are dominated by the ionised gas emission, while starting from 100\GHz\ an increasing contribution from thermal dust emission is evident, with a ring structure very similar to the main ring observed at optical and infrared wavelengths (Fig.\,\ref{HELIX}). For comparison, the 60 and 100\um\ IRIS maps are also shown in Fig.\,\ref{HELIX_IRIS}, where the same overall morphology as observed in the \Planck\ maps is evident. Apart from the well known structures associated with the evolved PN, diffuse emission is also present up to one degree from the central object. Such structures are evident from 100\GHz\ to 353\GHz\ and are very probably due to dust. To model the SED of the Helix, we followed the same procedure as described in Sec. \ref{sec:properties} for the other sources. Aperture photometry was performed, as described in Sect. \ref{sec:sed}, on both \Planck\ and IRIS maps. A colour correction was also applied to the \Planck\ fluxes. Only IRIS maps at 60 and 100\um\ were considered, as they are free from contamination from line emission \citep{Speck2002} and should trace only the dust continuum. The size of the radio source was set to 400\arcs\ (6\parcm7), as derived from the cuts performed across the radio source by \citet{Rodriguez2002}. All the relevant parameters of the free-free and {\tt DUSTY} fits are summarized in Tables~\ref{tab:ff} and \ref{tab:dust}. Encouragingly, our minimizing procedure provides an internal radius of the radio source (about 240\arcs) that is in good agreement with \citet{Rodriguez2002}. The results from our modelling of both free-free and thermal dust emission are shown in Fig.\,\ref{sp:Helix}, where a clear thermal dust contribution is superimposed on a typical optically-thin free-free spectrum. \Planck\ data agree well with data from the \WMAP\ 7-year catalogue and with the 31\,GHz data point from \citet{Casassus2004}. However, the \Planck\ data are fundamental to constrain the free-free and dust contributions, filling for the first time the observational range between 10 and $10^{3}$\GHz\ where the lack of measurements has led to many speculations \citep{Casassus2004}. As in the case of CRL\,618, the physical properties of the Helix, as derived from our fit, stand out with respect to the other targets. This is a consequence of its evolutionary stage, since the Helix is the most evolved PN of the sample; it is very extended (low $n_{\mathrm e}$), almost completely ionised (a high value of $M_\mathrm{ion}$), and embedded in an envelope of mostly low temperature dust.

\subsection{The morphology of the dust component} 
Due to the large angular extent on the sky (up to 14\arcm\ radius) and the angular resolution of \Planck\ at HFI frequencies (down to 5\arcm), it is possible to attempt a morphological study of the outer structures observed in the Helix directly from the \Planck\ maps.

The dust component of the Helix was already observed in the mid-IR and FIR using \ISO\ and \Spitzer\ \citep{cox_1998,Speck2002,Hora2004}. One of the major outcomes from such studies was the presence of a very peculiar size distribution of the dust grains, with a lack of small molecular-sized dust particles, which were probably destroyed by the strong radiation field of the central star \citep{cox_1998}. 
As a consequence, the \ISO\ and IRAC \citep{fazio_etal2004} maps are mainly tracing H$_{2}$ and fine-structure atomic lines.
The only information on the spatial distribution of the dust continuum comes from linear cuts performed on 90 and 160\um\ ISOPHOT images by \citet{Speck2002} at one particular position of the nebula (PA=155$\deg$), with an angular resolution (FWHM) of 44\parcs5 and 97\parcs4, respectively. The HFI maps, even with their low angular resolution, allow us for the first time to analyse the full spatial distribution of the dust component and to perform a comparative study of the different components coexisting in the nebula (ionised gas, thermal dust, and molecular gas) for a proper modelling of its physics.

\begin{figure}
\centering
\includegraphics[width=88mm]{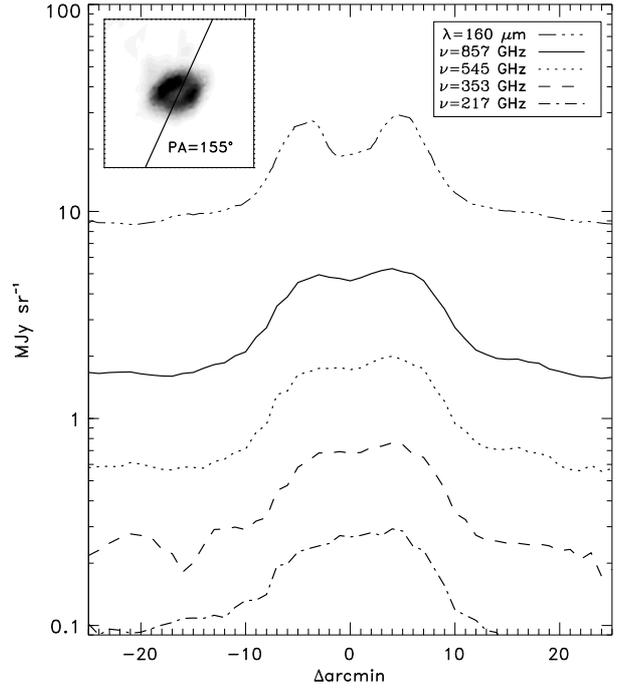}
\caption{Profiles obtained from the \Planck\ maps of the Helix at 217 (dash dot line), 353 (dashed line), 545 (dotted line), and 857\GHz\  (continuous line), with cuts at the same PA as in \citet{Speck2002}. The direction along which the cuts were performed is indicated in the small inset on the top left (superimposed on the 857\GHz\ \Planck\ map, in equatorial coordinates). The \Planck\ profiles are compared to the 160\um\ profile reported in \citet{Speck2002} (triple dot dashed line).}
\label{slice155}
\end{figure}

We first check if the results from the \Planck\ maps are consistent with the \ISO\ linear cuts \citep{Speck2002}. To do this, we performed cuts on the 217, 353, 545, and 857\GHz\ \Planck\ maps across the entire dust nebula at the same position angle (PA) of 155$\deg$  east to north, equatorial, as in \citet{Speck2002}. The 217, 353, 545, and 857\GHz\ profiles are shown in Fig.\,\ref{slice155}, together with the 160\um\ profile adapted from \citet{Speck2002}. The overall dust distribution agrees well with the ISOPHOT linear scan, indicating that in both \Planck\ and ISOPHOT profiles there is evidence of low-brightness dust emission out to a distance of about 18\arcm\ from the central star. The lack of emission from the centre reported by \citet{Speck2002} in the 160\um\ data, which they interpreted as evidence of a lack of small dust grains, is less evident in the \Planck\ cut at 857\GHz, but this can be ascribed to the lower angular resolution (1\parcm6 versus about 5\arcm).

\begin{figure}
\centering
\includegraphics[width=88mm]{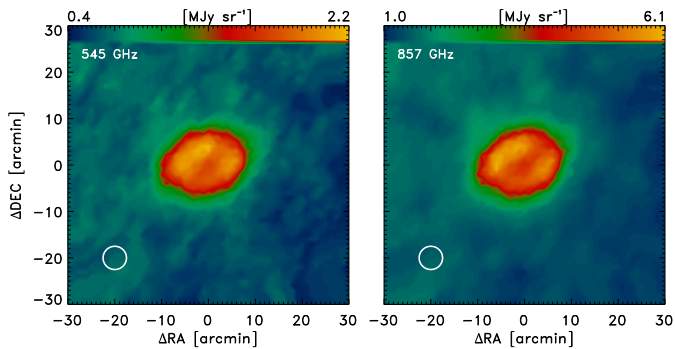}
\caption{Close-ups of the \Planck\ maps of the Helix (in equatorial coordinates) at 545 (left) and 857\GHz\ (right).} 
\label{helix_zoom}
\end{figure}

Close-ups of the 545 and 857\GHz\ \Planck\ maps are shown in Fig.\,\ref{helix_zoom}. The high-frequency maps have a quite similar morphology, indicating that they are tracing the same dust component. In particular, the 857\GHz\ image provides more details on the dust distribution, which can be summarized as follows.
\begin{enumerate}[a)]
\item An inner, low-brightness region extending up to 5\arcm.
\item A main ring of emission extending up to about 10\arcm\ from the central object. Here the dust is not uniformly distributed, but it appears more concentrated in the north-east than in the south-west part.
\item An extended, low-brightness structure extending up to 20\arcm\ from the central object in the north-east part of the nebula. Such extended structure is also visible at 545\GHz, but merges with the ISM at other frequencies.
\end{enumerate}

\begin{figure}
\centering
\includegraphics[width=88mm]{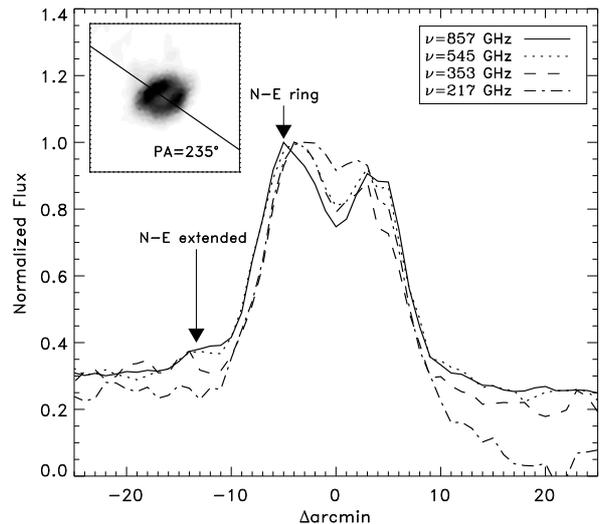}
\caption{Profiles of linear scans performed in the \Planck\ maps at 
217 (dash dot line), 353 (dashed line), 545 (dotted line) and 857\GHz\  (continuous line)
normalized to the 217\GHz\ profile, to better compare them. The direction along which the cuts have been performed is indicated in the small inset on the top left. It is superimposed to the 857\GHz\ \Planck\ map (PA=235$\deg$) and has been chosen in order to include the extended structure.}
\label{slice235}
\end{figure}

All these structures can also be seen in Fig.\,\ref{slice235}, which shows cuts of the 217, 353, 545, and 857\GHz\ maps along PA=235$\deg$, 
  where the minus side of the cuts correspond to the left-hand side of the inset map. While the presence of an inner cavity was already pointed out in \ISO\ observations and interpreted in terms of small dust grains being rapidly destroyed by the strong UV field of the central star \citep{cox_1998,Speck2002}, the presence of extended low-brightness structures is a unique result of the present observations. A hint of its presence was pointed out from linear cuts performed with ISOPHOT by \citet{Speck2002}, but only \Planck\ has provided us with the first (even though at low-resolution) mapping of this dusty structure, revealing its axisymmetric morphology, which is more extended toward the north-east.
Such extended low-brightness dusty emission seems spatially associated with the north-east outer arc clearly visible in the optical and ultraviolet images. Interestingly, \citet{Zhang2012} recently reported the discovery of a mid-IR halo around the Helix observed with the {\it Wide-field Infrared Survey Explorer} ($\WISE$) at 12\um, whose dimensions and morphology are very similar to those observed in the extended \Planck\ structure. As the mid-IR halo was not detected at  22\um\  and marginally detected by \WISE\ at 3.4\um\  and 4.6\um, \citet{Zhang2012} concluded that its emission peaks at 12\um\ and attributed the emission to a warm ($T_\mathrm{dust} $ about 300\,K) dust component. However, the broader emission observed in the HFI channel implies that there is also a contribution from colder dust.

It is not possible to directly compare our results from the {\tt DUSTY} fitting with the morphological information provided by \Planck\ maps. In fact, while there is an overall agreement between the inner radius for dust distribution, the value of the total extension of the dusty CSE provided by {\tt DUSTY} corresponds to regions where the dust reaches the temperature of the ISM, which is assumed to be 12\,K in the modelling. The thermal emission from such cold dust will be very weak at the \Planck\ and \IRAS\ frequencies, which are mainly tracing warmer dust.

\begin{figure}
\centering
\includegraphics[width=88mm]{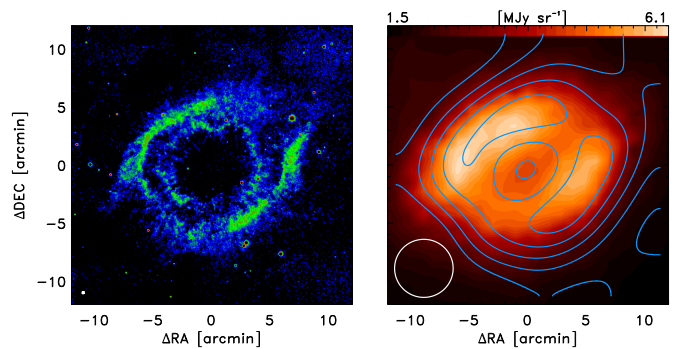}
\caption{{\it Left}: {\bf H$_{2}$} map of the Helix \citep{Hora2004}. {\it Right}: the comparison between the H$_{2}$ map (contours) and the \Planck\ 857\GHz\ map. Only the main-ring emission of the dust component has been reproduced. The H$_{2}$ map has been smoothed to the angular resolution of the \Planck\ map, indicated in the left-bottom corner.}
\label{h2}
\end{figure}

It is interesting to analyse how the dust component observed by \Planck\ is morphologically related to the molecular and ionised gas component. To do so, we compare the \Planck\ 857\GHz\ map to the ground-based near-IR molecular Hydrogen (2.12\um) image from \citet{Hora2004} and to the radio map at 1.4\GHz\ from the NVSS. In the H$_{2}$ map, the well-known double ring structure is clearly seen (see Fig.\,\ref{h2}) and is very similar to that observed in the optical image. Before making a direct comparison of the two maps, we subtracted the very bright emission from field stars in the H$_{2}$ image and then convolved it with the \Planck\ beam at 857\GHz. The comparison between the two components is shown in the right-hand panel of Fig.\,\ref{h2}.

The molecular component is spatially confined within the dust component (only the main ring of dust emission has been reproduced in the image). However, we cannot exclude the presence of H$_{2}$ beyond the dust emitting region, since at this distance from the central star H$_{2}$ is not excited, and therefore, even if present, it cannot be detected \citep{Aleman2011}.

\begin{figure*}
\includegraphics[width=180mm]{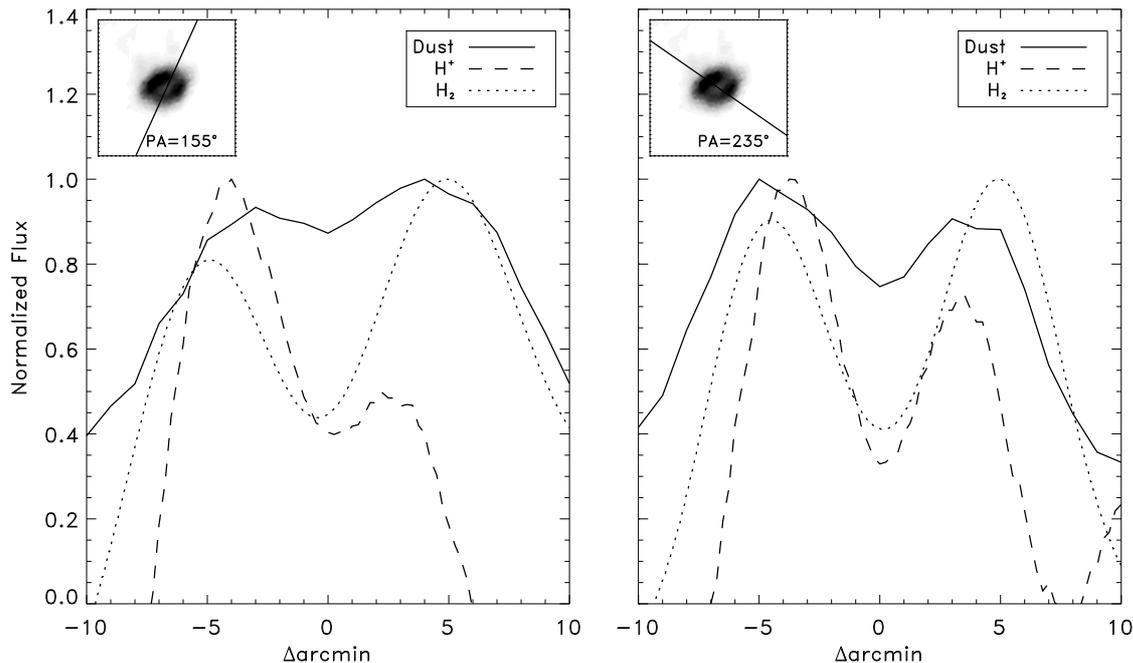}
\caption{Profiles of the brightness of the dust 
(\Planck\ at 857\GHz\ -- continuous line), radio (H+, NVSS at 1.4\GHz\ -- dashed line) and H$_{2}$ (dotted line)
components obtained at PA=155$\deg$ (left) and 235$\deg$ (right). Profiles have been normalized to the Helix peak of each maps.}
\label{radio}
\end{figure*}

In order to study the morphology of the ionised gas, we retrieved a radio image of the Helix at 1.4\GHz\ from the NVSS web archive\footnote{\url{http://www.cv.nrao.edu/nvss/postage.shtml}}. The radio morphology is the same as in \citet{Rodriguez2002}, where there are several point sources in the field that are not related to the nebula, but are probably extragalactic sources. To better determine the relative distributions of each component in the nebula, profiles have been obtained with cuts performed at two set position angles in the radio (NVSS), dust (\Planck), and H$_{2}$ maps (Fig. \ref{radio}). The maps have first been smoothed to the angular resolution of the \Planck\ 857\GHz\ map. The radio map traces the inner ring, where most of the radio emission comes from, surrounded by both the molecular (H$_{2}$) and dust (\Planck\ maps) components.

NVSS observations, even though taken at 1.4\GHz\ with the VLA in its most compact configuration (typical synthesized beam of the order of 45\arcs), are not sensitive to structures larger than around 15\arcm. Therefore, in the NVSS map, very extended low-brightness structures, such as those observed in $H_\alpha$ and reported by \citet{Speck2002}, might be resolved out. \citet{Rodriguez2002} pointed out that the ionised gas, as traced by the continuum radio observations, is surrounded by \hone\ emission (21cm line) as indicated by the observations carried out with a similar instrumental configuration. The \hone\ distribution well reproduces the typical helical structure of the molecular envelope, with a morphological similarity with the H$_{2}$ emission, and is actually delineating the outer parts of the bright ring-like H$_{2}$ region. The ring-like structure of the inner ionised gas versus the more extended morphology of the molecular envelope is also supported by the images obtained using ISOCAM with the LW2 filter (centred at 6.9\um) and the LW3 filter (centred at 15\um) by \citet{cox_1998}. Again, while the former traces the H$_{2}$ emission and thus delineates the molecular envelope morphology with the classical helical structure, the  latter is dominated by the  [\ion{Ne}{iii}] emission and thus traces the ionised gas, clearly showing a more compact, rounder morphology. The interplay between the dust and ionised components is likely to evolve
with the nebular expansion \citep{chu}. At the angular
resolution of Planck, our data indicate that  the ionised material is concentrated within the dusty nebula, closer to the central star.

\section{Conclusions} \label{sec:conclusions}

The remarkable frequency coverage of \Planck\ has allowed us to collect important data for a small sample of Galactic PNe. These PNe, being among the brightest of such sources, are the best studied in our Galaxy. Nevertheless, a comprehensive picture of their CSEs in both their main components, i.e., ionised gas and dust, has been provided for the first time.

In particular, the evaluation of the emission from ionised gas (free-free) allows us to constrain the thermal dust emission. The modelling of the SEDs provides us with good estimates of the physical parameters of the CSEs. From our modelling, the density, spatial distribution and total mass of ionised gas have been derived, as well as the internal radius, the extent of the dusty envelope and its mass content. 

One interesting result is that, in general in the studied targets, dust and ionised gas appear to be partially co-spatial, which implies the existence of some kind of shielding mechanism to allow the dust to form, or at least to survive, in the harsh environment created by the strong UV radiation field of the central object. Such shielding can be provided by a dusty disk or a molecular torus, as pointed out by interferometric observations of young, bi-polar PNe, i.e., CRL\,618, or can be in the form of clumps of material, as observed in older objects such as the Helix or NGC\,6720. These shielding structures must be very common and may well be debris from material existing before the nebula was formed.
 
Thanks to the multi-frequency \Planck\ measurements, the SED of the variable source CRL\,618 was modelled for the first time, including both ionised gas and thermal dust contributions. This allowed us to derive important physical parameters of the CSE without variability effects.
 
Finally, in the case of the Helix, \Planck\ maps enable us to perform a morphological study of the extended circumstellar material associated with the evolved PN. The dust emission was fully mapped for the first time and three main components were found, including an extended structure already seen in $H_\alpha$ observations, probably related to a region where the slow expanding envelope interacts with the surrounding ISM. A comparison was also performed between the morphology of the dust component as traced by \Planck, of the molecular gas traced by H$_{2}$ near-IR observations, and of the ionised gas traced by radio (1.4\GHz) observations. While the dust and H$_{2}$ share a comparable morphology, the ionised gas appears more concentrated in the inner ring of the nebula.

Now that Planck has robustly determined the large-scale emission of this PN sample over a factor of 30 in wavelength,
 there is a firm basis within which to carry out detailed follow-up studies of the higher spatial and spectral-resolution
  properties. In particular, the results presented in this paper will provide us with a useful  
  framework within which to plan future projects with ALMA, focusing on detailed morphological studies 
  of both the ionised and dusty components.

\begin{table*}[tmb]
\appendix
\section{}
\begingroup
\newdimen\tblskip \tblskip=5pt
\caption{Additional informations on the mm and sub-mm ancillary data used to make the SEDs.}
\label{tab:info}
\nointerlineskip
\vskip -3mm
\footnotesize
\setbox\tablebox=\vbox{
 \newdimen\digitwidth 
 \setbox0=\hbox{\rm 0} 
 \digitwidth=\wd0 
 \catcode`*=\active 
 \def*{\kern\digitwidth}
 \newdimen\signwidth 
 \setbox0=\hbox{+} 
 \signwidth=\wd0 
 \catcode`!=\active 
 \def!{\kern\signwidth}
 \halign{\hbox to 2in{#\leaderfil}\tabskip .5em&
\hfil#\hfil&
 \hfil#\hfil&
 \hfil#\hfil&
 \hfil#\hfil\tabskip 0pt\cr
 \noalign{\doubleline\vskip 2pt}
Facility &Bands &Beam or Aperture Size &notes &References \cr
\omit & &[arcsec] & & \cr
\noalign{\vskip 4pt\hrule\vskip 6pt}


KAO & 36 / 53 / 61 / 131 [$\mu$m] & 28 / 50  & \s beam & \citealp{Telesco1977} **\cr
Hale telescope            &  1 [mm]                                    & 55        &\s beam & \citealp{Elias1978}**\cr
KAO & 37 / 52 / 70 / 108 [$\mu$m] & 20 / 27 / 55 &\s beam & \citealp{Moseley1980}**\cr
NRAO Millimeter-Wave Telescope & 90 / 150 [GHz] & 78 / 52 & \s beam & \citealp{Ulich1981}**\cr
UKIRT & 370 / 780 / 1090 [$\mu$m]  & 40 -- 64    &\s beam & \citealp{Gee1984}**\cr
IRAM 30-m & 90 [GHz] &26  & \s beam & \citealp{steppe_etal88}**\cr
JCMT &450 / 800 / 1100 / 2000 [$\mu$m]  &18 / 19 / 26  & \s aperture size & \citealp{Hoare1992}**\cr
JCMT & 350 -- 2000 [$\mu$m]  & $\approx 16$ -- 27  & \s beam & \citealp{knapp_etal93}**\cr 
IRAM 30-m & 250 [GHz] & $\approx 12$ & \s beam &  \citealp{altenhoff_etal94}**\cr
JCMT & 350 -- 2000 [$\mu$m] & $\approx 16$ -- 27   & \s beam & \citealp{sandelll94}**\cr
ISOPHOT (mf)& 7.3 -- 200 [$\mu$m] & 23 -- 184   & \s aperture size & \citealp{Klaas2006}**\cr
SEST & 99 / 147 / 230 / 250 [GHz] & 50 /  34 / 24  & \s beam & \citealp{casassus_etal07}**\cr
%
\noalign{\vskip 3pt\hrule\vskip 4pt}
}}
\endPlancktable
\endgroup
\end{table*}


\begin{acknowledgements} 
The development of \Planck\ has been supported by: ESA; CNES and CNRS/INSU-IN2P3-INP (France); ASI, CNR, and INAF (Italy); NASA and DoE (USA); STFC and UKSA (UK); CSIC, MICINN, JA and RES (Spain); Tekes, AoF and CSC (Finland); DLR and MPG (Germany); CSA (Canada); DTU Space (Denmark); SER/SSO (Switzerland); RCN (Norway); SFI (Ireland); FCT/MCTES (Portugal); and PRACE (EU). A description of the Planck Collaboration and a list of its members, including the technical or scientific activities in which they have been involved, can be found at \url{http://www.sciops.esa.int/index.php?project=planck&page=Planck_Collaboration}
The National Radio Astronomy Observatory is a facility of the National Science Foundation operated under cooperative agreement by Associated Universities, Inc.
L. Cerrigone acknowledges funding from the Spanish Consejo Superior de Investigaciones Cient{\'{\i}}ficas through a JAE-Doc research contract, co-funded by the European Social Fund.  L.~C. thanks the Spanish MICINN for funding support through grants AYA2009-07304 and CSD2009-00038. 
\end{acknowledgements}

\bibliographystyle{aa} 
\bibliography{Planck_bib,PN_Planck} 

\raggedright

\end{document}